\documentstyle[preprint,aps]{revtex}
\tightenlines
\begin{document}
\draft
\title{ Theory of weak continuous measurements \\
in a strongly driven quantum bit.}
\author{Anatoly Yu. Smirnov }
\address{ D-Wave Systems Inc. 320-1985 West Broadway,\\
 Vancouver, B.C. V6J 4Y3, Canada }
\date{\today}
\maketitle

\begin{abstract}
{ Continuous spectroscopic measurements of a strongly driven superconducting qubit by means of a high-quality tank circuit (a linear detector) are under study. Output functions of the detector, namely, a spectrum of voltage fluctuations and an impedance, are expressed in terms of the qubit spectrum and magnetic susceptibility. The nonequilibrium spectrum of the current fluctuations in the qubit loop and the linear response function of the driven qubit coupled to a heat bath are calculated with Bloch-Redfield and rotating wave approximations. Backaction effects of the qubit on the tank and the tank on the qubit are analyzed quantitatively. We show that the  voltage spectrum of the tank provides detailed information about a frequency and a decay rate of Rabi oscillations in the qubit.
It is found that both an efficiency of spectroscopic measurement and measurement-induced decoherence of the qubit demonstrate a resonant behaviour as the Rabi frequency approaches the resonant frequency of the tank.
We determine conditions when the spectroscopic observation of the Rabi oscillations in the flux qubit with the tank circuit can be considered as a weak continuous quantum measurement.
}
\end{abstract}

 \maketitle

\section{Introduction}

Quantum electrical engineering treats electrical circuits as macroscopic quantum systems. Some of these systems are approximately characterized by two states, and, therefore, can be considered as prototypes of a quantum bit, a main element of quantum computers. Recently, an existence of two quantum states has been proven experimentally in electrical circuits based on Josephson junctions \cite{Makhlin1,Nakamura1,Wal1,Yu1,Vion1,Chiorescu1,Il'ichev1}; in so doing coherent oscillations between macroscopically different states have been measured both in a free-evolution regime (quantum beatings) and in the presence of resonant driving field (Rabi oscillations). In the majority of the experiments a detector, for example, a dc-SQUID, was strongly coupled to the qubit during the measurement that resulted in a fast collapse of the circuit into one of its eigenstates.
For mapping a whole evolution of the qubit the measurements were repeated thousands times  with the same conditions to gather a statistical ensemble. An alternative procedure has been employed in an experiment \cite{Il'ichev1} where continuous measurements of Rabi oscillations in a three-Josephson-junction (3JJ) flux qubit have been performed. In this experiment the qubit is inductively coupled to a high-quality tank circuit that serves as a linear detector. Rabi oscillations reveal themselves in a spectrum of voltage fluctuations in the tank as the Rabi frequency of the qubit $\Omega_R$ passes through the resonant frequency of the tank $\omega_T$.  It should be noted that the approach undertaken in Ref. \cite{Il'ichev1} can be considered as an experimental realization of weak continuous measurements that have been studied theoretically in Refs. \cite{Averin1,Averin2,Pilgram1,Clerk1,Korotkov1,Brink1}. Previously an effect of which-path detector on electron dephasing in a double-path interferometer was demonstrated experimentally in Refs.\cite{Buks1,Sprinzak1}.
Weak coupling between the qubit and the detector (tank) does not yield complete information about the state of the qubit before and during the measurement while not introducing severe decoherence into the quantum system. Because of the last reason the qubit can be monitored continuously with an extraction of useful spectroscopic information about quantum processes going on in the system. Among other things, a Rabi frequency  ($\Omega_R/2\pi = 6.284 $MHz) and a life-time of Rabi oscillations ($\tau_{Rabi} = 2.5 \mu s$)
 in the strongly driven qubit have been measured in Ref. \cite{Il'ichev1}. We interpret the qubit as a strongly driven system if the frequency of Rabi oscillations, $\Omega_R$, exceeds significantly the damping rate of the qubit, $\Gamma: \Omega_R \gg \Gamma,$ while being much less than the frequency of quantum beatings.
Here we have to mention some distinctions between the theoretical analysis of the weak quantum  measurements given in Refs.\cite{Averin1,Averin2} and the experimental implementation \cite{Il'ichev1}. An assumption made in the theory that a characteristic time of the detector is much shorter than the period of measured oscillations is not valid for the tank (detector) with the resonant frequency $\omega_T$ of order of the Rabi frequency $\Omega_R$ which is measured in the experiment. This discrepancy is cleared up with a replacement of a local-in-time response coefficient of the detector by a nonlocal response function of the tank. In the  output spectrum we will have now a product of the qubit spectrum and a Lorentzian that describes a susceptibility of the tank (a Fourier transform of the tank response function).
The next distinction is related to the fact that the analysis performed in Refs. \cite{Averin1,Averin2} is aimed at the continuous measurements of quantum beatings which are determined by the equilibrium spectrum of the qubit. This spectrum has a peak at the frequency of quantum oscillations between degenerate states of the qubit. By contrast, the experimental setup deployed in Ref.\cite{Il'ichev1} is dealing with a strongly nonequilibrium situation when the measured quantum system is driven by the external microwave field. In this case the output spectrum of voltage fluctuations in the tank (detector) depends on the nonequilibrium current (flux) noise in the qubit loop and has a peak at the Rabi frequency which is much lower than the frequency of quantum beatings but much higher than the decay rate of the qubit. Emission and absorption spectra of the driven atoms (two-level systems) have been investigated by Mollow \cite{Mollow1} in the Markov approximation. It was shown that in the presence of a strong driving field a single line-shape function of the atom bifurcates into two Lorentzians shifted from the usual resonant frequency of the atom by the Rabi frequency which is proportional to the amplitude of the driving field. The signals from the atom can be detected by means of high-frequency spectroscopic devices only. The two-level system  measured in Ref.\cite{Il'ichev1} produces also a low-frequency output that can be associated with biasing of the qubit from the degeneracy point \cite{Greenberg1}. We suppose as well that relaxation and decoherence rates can not be introduced into the qubit equations phenomenologically as it was done by Mollow for the atomic system. The results of the experiment \cite{Chiorescu1} are indicative of a pronounced dependence of the qubit damping rates on the amplitude of the driving force. In particular, a decay time of Rabi oscillations, $\tau_{Rabi} \simeq 150 ns$, measured in Ref. \cite{Chiorescu1}, is significantly different both from a dephasing time ($\tau_{\varphi} = 20 ns$) and from a relaxation time ($\tau_{relax} = 900 ns$) of the undriven flux qubit. It is shown theoretically \cite{Smirnov1} that well-known formulas for the dephasing and relaxation times \cite{Slichter1,Grifoni1} are no longer valid for the strongly driven qubit. In this case  relaxation and dephasing are mixed and determined by the spectrum of the heat bath fluctuations $S(\omega)$ taken at the Rabi frequency as well as at combinations of the energy splitting of the qubit, $ \omega_c = \sqrt{\Delta^2 + \varepsilon^2}$, and the Rabi frequency $\Omega_R$: $ \omega_c \pm \Omega_R.$ Here $\Delta $ and $\varepsilon$ are a tunneling rate of the qubit and a bias, respectively. Besides that, the driven qubit is no longer in thermodynamic equilibrium with the heat bath as evidenced by zero value of a steady-state population difference between the qubit energy levels \cite{Smirnov1}. Because of this, we can not resort to the Callen-Welton fluctuation-dissipation theorem \cite{CallenWelton} to find the spectrum of qubit fluctuations.

This paper is devoted to a detailed consideration of nonequilibrium fluctuations and decoherence  in a strongly driven qubit coupled to a linear detector (a high-quality tank). To accomplish these ends we apply a formalism of quantum stochastic equations proposed and developed in Refs.\cite{Efremov1,Smirnov2,Mourokh1}. Our quantitative analysis is motivated by the recent experiments \cite{Il'ichev1,Chiorescu1} and based on an assumption of a weak interaction between the qubit and the detector. This assumption can fail near the point of exact resonance between the qubit oscillations and electromagnetic oscillations in the tank, in particular, at the point  $\Omega_R = \omega_T.$ We define conditions whereby the measurements of Rabi oscillations in the flux  qubit performed in Ref. \cite{Il'ichev1} with a high-quality LC circuit (a tank)  fall into the  category of weak quantum measurements \cite{Averin1,Averin2}.  With this aim in mind we calculate a spectrum of voltage fluctuations in the tank as well as a contribution of the detector into decoherence rate of the qubit.
The experimental set-up implemented in Ref.\cite{Il'ichev1} (so called impedance measurement technique) can also monitor an effective impedance of the system "qubit+tank" by applying a small ac current, $I_{bias}$, to the tank with a subsequent measurement of an angle between the tank voltage and the ac current \cite{Il'ichev2,Greenberg2}.
This angle is determined by the magnetic  susceptibility of the qubit provided that the frequency of the ac current coincides with the resonant frequency of the tank. Here we calculate the magnetic susceptibility of the strongly driven qubit together with its decay rates taking into account detuning between the high-frequency source and the qubit.

The present paper is organized as follows. Dynamics and fluctuations in the linear detector (a tank circuit) coupled to the qubit are under study in Sec.II. We derive expressions for the averaged voltage in the tank, for the angle between the voltage and ac current, as well as for the spectrum of voltage fluctuations in terms of the magnetic susceptibility of the qubit and the spectrum of qubit fluctuations. In Sec.III we derive Heisenberg-Langevin equations for the driven qubit interacting with the tank and with its internal heat bath which is responsible for the qubit decoherence in the absence of the tank. These equations are subsequently simplified using a rotating wave approximation for the qubit that is weakly coupled to its environment, i.e. to   the tank and to the internal bath. The decay rates of the qubit depending on the amplitude of the driving force and on its detuning from energy splitting of the qubit are derived in Sec.III. A dissipative evolution of a probability to find the qubit in the excited state as well as an evolution of an averaged current in the qubit loop are considered in Sec.IV. In the same section linear response functions of the driven two-level system are calculated together with parameters which are required for the impedance measurement technique (IMT).
The nonequilibrium spectrum of qubit fluctuations are found in Sec.V for the case of zero detuning.
The output of the linear detector, namely, the spectrum of voltage fluctuations in the tank, and the contribution of the measuring device into qubit decoherence are presented also in Sec.V. In Appendices we outline our approach to the theory of open quantum system (Appendix A) and explain in more detail a derivation of collision terms (Appendix B) and spectra of fluctuation forces (Appendix C).

\section{ LC-circuit inductively coupled to the qubit}

We consider a 3JJ flux qubit \cite{Mooij1,Orlando1} driven by a strong high-frequency field and inductively coupled to a tank circuit.
This coupling is proportional to the coefficient of qubit-tank mutual inductance $ k\sqrt{L_q L_T}$ as well as to the product of currents in the tank $I_T$ and in the qubit loop $I_q$. Here $k$ is a dimensionless coupling parameter, $L_q$ and $L_T$ are the inductances of the qubit loop and the tank, respectively, $C_T$ is a tank capacitance.  In the quantum case an operator of the qubit current is determined by $\sigma_z$ matrix, $\hat{I}_q = I_q \sigma_z$, that corresponds to two directions of the persistent current in the qubit loop. An operator of the current in the LC-circuit (tank) is defined in terms of creation-annihilation operators of photons in the tank having a resonance frequency $\omega_T = 1/\sqrt{L_T C_T}:$ $\hat{I}_T = \sqrt{\hbar \omega_T/2 L_T}(a + a^+), [a,a^+]_- = 1. $  For the Hamiltonian of the total "qubit-tank" system we have the expression
\begin{equation}
H = {\Delta\over 2} \sigma_x + {\varepsilon \over 2} \sigma_z - \sigma_z F \cos\omega_0t -  \sigma_z ( Q_0 + f  + \lambda \hat{I}_T) + H_T + H_{qB}
\end{equation}
where
$\Delta$ is a tunneling rate of the qubit,  $\varepsilon$ is a bias, $\lambda = kI_q \sqrt{L_q L_T}$ is the coupling coefficient between the qubit and the tank. An operator $Q_0$ describes an internal dissipative environment of the qubit (without the tank), $ H_{qB} $ is a Hamiltonian of this heat bath. The heat bath $Q_0$ corresponds to all sources of external flux noise which are additional to the noise created by the tank \cite{Tian1,Wilhelm1}.
In particular, the flux qubit is coupled to nuclear and impurity spins that can contribute to its decoherence and dephasing \cite{Stamp1,Rose1}.
We also introduce here $f(t)$, a small external force that is required for a calculation of magnetic susceptibility and, in particular, an absorption spectrum of the qubit.
In its turn, the tank driven by a bias current $ I_{bias} $ is characterized by the Hamiltonian $H_T$,
\begin{equation}
H_T = \hbar \omega_T (a^+a + 1/2) - (a + a^+)Q_b  - L_T \hat{I}_T I_{bias} + H_{TB}
\end{equation}
$Q_b$ is a variable of another heat bath which directly interacts with the tank. That heat bath, having a free Hamiltonian $H_{TB}$,
 is responsible for finite life time of the photons, $\gamma_T^{-1},$ as well as for a finite quality factor of the tank, $Q_T = \omega_T /2\gamma_T.$

Operators of the tank current, $\hat{I}_T,$ and tank voltage,$\hat{V}_T = i\sqrt{\hbar \omega_T/2C_T} (a^{+} - a),$ obey the equations: $\dot{\hat{I}}_T = \hat{V}_T/L_T, $
\begin{equation}
\left( {d^2\over dt^2} + \omega_T^2\right) \hat{V}_T = \sqrt{2\omega_T \over C_T}\dot{Q}_b + \lambda \omega_T^2 \dot {\sigma}_z + {1\over C_T} \dot{I}_{bias}.
\end{equation}
From here on we put $\hbar = 1,$ and $ k_B = 1.$
For a small coupling between the tank and its own bath and/or for the case of Gaussian fluctuations of free variables of this heat bath, $Q_b^{(0)}$, a total operator $Q_b(t)$ has the form
\begin{equation}
Q_b(t) = Q_b^{(0)}(t) + \sqrt{2L_T/\omega_T} \int dt_1 \varphi_b(t,t_1) \hat{I}_T(t_1),
\end{equation}
with
\begin{equation}
\varphi_b(t,t_1) = \langle i [Q_b^{(0)}(t),Q_b^{(0)}(t_1)]_-\rangle \theta (t-t_1)
\end{equation}
being a linear response function of the bath.
Here $\theta(\tau)$ is the Heaviside step function, and $ \langle Q_b^{(0)} \rangle = 0.$

To characterize this heat bath thoroughly we introduce also a correlation function of the unperturbed variables $Q_b^{(0)}$:
\begin{equation}
M_b(t,t_1) = \langle (1/2) [ Q_b^{(0)}(t),Q_b^{(0)}(t_1)]_+\rangle
\end{equation}
together with a corresponding spectral funcion $S_b(\omega )$  which represents a Fourier transform of $M_b(\tau )$.
In the case of Ohmic dissipation in the tank with a resistance $R_T$ the imaginary part of the susceptibility $\chi_b(\omega )$, corresponding to the response function $\varphi_b(t-t_1)$, is proportional to the frequency $\omega $:
$\chi^{\prime\prime}_b(\omega ) = (\gamma_T /2\omega_T) \omega,$ and to the line width of the tank $\gamma_T = 1/(R_T C_T)$ with $\varphi_b(\tau) = - (\gamma_T /2\omega_T)(d/d\tau ){\delta}(\tau ).$  According to the fluctuation-dissipation theorem for the spectrum $S_b(\omega ) $ we obtain:
\begin{equation}
S_b(\omega ) = (\gamma_T /2\omega_T) \omega \coth (\omega /2T),
\end{equation}
where $T$ is an equilibrium temperature of the heat bath coupled to the tank. We suppose that this initial temperature is equal to the equilibrium temperature of the bath interacting with the qubit.

In the presence of a time-dependent bias current, $I_{bias}(t)$, the equation (3) for the operator of the tank voltage can be rewritten in the form of a simple stochastic equation
\begin{equation}
\left( {d^2\over dt^2} + \gamma_T {d \over dt} + \omega_T^2\right) \hat{V}_T = \sqrt{2\omega_T \over C_T}\dot{Q}_b^{(0)} + \lambda \omega_T^2 \dot {\sigma}_z + {1\over C_T} \dot{I}_{bias}.
\end{equation}
A qubit operator $\dot {\sigma}_z  $ in the right-hand side of Eq.(8) functionally depends on the tank voltage $ \hat{V}_T. $ For a small qubit-tank interaction this dependence is approximately described by the formula:
\begin{equation}
 \dot{\sigma}_z  = \dot{\sigma}_{z,0} + {\lambda \over L_T} \int dt_1 \langle {\delta \sigma_z(t) \over \delta f(t_1)}\rangle \hat{V}_T(t_1).
\end{equation}
Here we take into account the relation $(\delta /\delta I_T) = \lambda (\delta /\delta f), $ as well as the equation: $\dot{\hat{I}}_T = \hat{V}_T/L_T. $ Fluctuations of the term
$ \dot{\sigma}_{z,0}, \langle\dot{\sigma}_{z,0}\rangle = 0, $ are determined by the internal bath of the qubit, $Q_0$, only, so that the operator $\dot{\sigma}_{z,0}$ has no correlations with the heat bath $Q_b$ coupled directly to the tank.
The functional derivative, $ \langle \delta \sigma_z(t)/\delta f(t')\rangle , $ has a magnetic susceptibility of the  qubit, $\chi_{zz}(\omega)$, as its Fourier transform:
\begin{equation}
\langle {\delta \sigma_z(t) \over \delta f(t')}\rangle= \int {d\omega \over 2 \pi}
e^{-i\omega(t-t')} \chi_{zz}(\omega).
\end{equation}
These functions describe a behaviour of the qubit current, $\langle \hat{I}_q \rangle = I_q \langle \sigma_z \rangle,$ induced by variations of the time-dependent external flux, which can be created by the tank, $f = k \sqrt{L_q I_q^2/L_T} \Phi_{rf}(t).$
Taking into account a qubit back-action on the tank  we obtain the following equation for the tank voltage:
\begin{eqnarray}
\int dt_1 \left[ \left( {d^2\over dt^2} + \gamma_T {d \over dt} + \omega_T^2\right)\delta(t - t_1) - {\lambda^2 \over L_T}\omega_T^2 \langle {\delta \sigma_z(t) \over \delta f(t_1)}\rangle \right]
 \hat{V}_T(t_1)  = \nonumber\\
 \sqrt{2\omega_T \over C_T}\dot{Q}_b^{(0)} + \lambda \omega_T^2
\dot {\sigma}_{z,0} + {1\over C_T} \dot{I}_{bias}.
\end{eqnarray}
The voltage $ \hat{V}_T(t) $, or, more precisely, its average value and correlation functions,
 serves as a meter in the process of qubit measurements.
Averaging of Eq.(11) over the initial state of the qubit and over the equilibrium fluctuations of all dissipative environments allows us to find the average tank voltage,
$\langle \hat{V}_T(t)\rangle = V_T \cos (\omega t + \Theta),$ induced by
the time-dependent bias current, $I_{bias}(t) = I_{ac}\cos \omega t.$ In the framework of the impedance measurement technique (IMT) \cite{Il'ichev2,Greenberg2} an imaginary part of the total impedance of the system "qubit+tank", defined by a voltage-current phase shift $\Theta$, is studied as a function of qubit parameters, such as a bias, etc..
Matching the coefficients before $e^{-i\omega t}$ in the averaged Eq.(11) gives the result:
\begin{equation}
V_T e^{-i\Theta} = - i \omega \left\{   \omega_T^2\left[ 1 - {\lambda^2 \over L_T} \chi_{zz}^{\prime }(\omega)\right] - \omega^2 - i \omega \left[ \gamma_T + {\lambda^2 \omega_T^2\over \omega L_T} \chi_{zz}^{\prime\prime}(\omega )\right] \right\}^{-1} {I_{ac} \over C_T}.
\end{equation}
Here $\chi_{zz}^{\prime }(\omega)$ and $\chi_{zz}^{\prime\prime }(\omega)$ are the real and imaginary parts of the qubit magnetic susceptibility (10).
At resonant conditions when the frequency of the bias current is exactly equal to the resonance frequency of the tank, $\omega = \omega_T,$ the amplitude of the voltage oscillations is determined by the equation
\begin{equation}
V_T = {I_{ac} \over C_T} \left\{ [  k^2 L_q I_q^2 \omega_T \chi_{zz}^{\prime}(\omega_T)]^2 + [ \gamma_T + k^2 L_q I_q^2 \omega_T \chi_{zz}^{\prime\prime}(\omega_T)]^2  \right\}^{-1/2}.
\end{equation}
For the voltage-current phase shift we obtain the expression
\begin{equation}
\tan \Theta = - k^2 L_qI_q^2  {\omega_T \chi_{zz}^{\prime}(\omega_T ) \over
\gamma_T + k^2 L_qI_q^2 \omega_T \chi_{zz}^{\prime\prime}(\omega_T ) }.
\end{equation}
Measurements of the angle between the average voltage in the tank and the bias current gives us an immediate information about the real part of the qubit magnetic susceptibility $\chi_{zz}^{\prime}(\omega_T ) $ taken at the resonance frequency of the tank $\omega_T.$  A sensitivity of the impedance measurement technique is adversely affected by a qubit contribution to the tank damping rate which is proportional to the imaginary part of the qubit susceptibility  $ \chi_{zz}^{\prime\prime}(\omega_T )$.

It follows from the stochastic part of Eq.(11) that a correlator of the voltage fluctuations (a cumulant function),
\begin{equation}
 M_V(t,t') = \langle (1/2)[\hat{V}_T(t),\hat{V}_T(t')]_+\rangle = \int {d\omega \over 2 \pi } e^{-i\omega (t-t')} S_V(\omega ),
\end{equation}
 satisfies the equation
\begin{eqnarray}
\int dt_1 \left[ \left( {d^2\over dt^2} + \gamma_T {d \over dt} + \omega_T^2\right)\delta(t - t_1) - {\lambda^2 \over L_T}\omega_T^2 \langle {\delta \sigma_z(t) \over \delta f(t_1)}\rangle \right] \times \nonumber\\
\int dt_2 \left[ \left( {d^2\over dt'^2} + \gamma_T {d \over dt'} + \omega_T^2\right)\delta(t' - t_2) - {\lambda^2 \over L_T} \omega_T^2 \langle {\delta \sigma_z(t') \over \delta f(t_2)}\rangle \right] M_V(t_1,t_2) = \nonumber\\
{d^2 \over dt dt'}\left\{ {2\omega_T \over C_T} M_b(t,t') + \lambda^2 \omega_T^4
\langle {1\over 2} \left[\sigma_{z,0}(t), \sigma_{z,0}(t')\right]_+\rangle  \right\}.
\end{eqnarray}
The total spectrum of the qubit, $ S_{zz}(\omega ) = S_{zz,0}(\omega ) + S_{zz,T}(\omega ) ,$
\begin{equation}
S_{zz}(\omega ) = \int d(t-t') e^{i\omega(t-t')} \langle {1\over 2} \left[\sigma_{z}(t), \sigma_{z}(t')\right]_+\rangle,
\end{equation}
incorporates a Fourier transform, $S_{zz,0}(\omega )$, of the correlator $\langle {1/2} \left[\sigma_{z,0}(t), \sigma_{z,0}(t')\right]_+\rangle,$  which is originated from qubit coupling to its internal heat bath, $Q_0,$
together with a contribution $ S_{zz,T}(\omega ),$ resulting from the qubit-tank interaction. This contribution has been built into the left-hand side of Eq.(16).

It follows from Eq.(16) that the total spectrum of voltage fluctuations in the tank $ S_V(\omega )$  (15) contains a contribution of the thermal noise $S_b(\omega ) $ (7) which is complemented by nonequilibrium noise generated by the qubit $S_{zz,0}(\omega ):$
\begin{equation}
S_{V}(\omega ) =  \omega^2 {\omega_T\over C_T} \times  { 2 S_b(\omega ) + k^2 L_q I_q^2  \omega_T   S_{zz,0}(\omega )   \over
(  \bar{\omega}_T^2  -  \omega^2 )^2 + \omega^2  \bar{\gamma}_T^2 }.
\end{equation}
Here $\bar{\omega}_T$ is a resonant frequency of the tank shifted in the presence of the qubit,
\begin{equation}
\bar{\omega}_T  = \omega_T\sqrt{ 1 - k^2 L_q I_q^2  \chi_{zz}^{\prime }(\omega_T)},
\end{equation}
and $\bar{\gamma}_T$ is a tank linewidth having regard to the qubit contribution to the tank damping,
\begin{equation}
 \bar{\gamma}_T =  \gamma_T + k^2 L_q I_q^2 \omega_T \chi_{zz}^{\prime\prime}(\omega_T )
\end{equation}
A frequency shift of the tank and a correction to the tank damping rate depend on the total susceptibility of the qubit, $\chi_{zz}(\omega )$, that should be calculated with consideration for all mechanisms of qubit dissipation.

\section{ Quantum Langevin equations.}

In this section we derive  Heisenberg-Langevin equations with a subsequent goal
of finding the nonequilibrium spectrum $ S_{zz}(\omega )$ of the qubit together with  its magnetic susceptibility $ \chi_{zz} (\omega ). $
To do that we consider  a two-state system (a quantum bit)  interacting with a heat bath $Q$ in the presence of a harmonic driving force $F(t) = F \cos \omega_0t.$
This heat bath incorporates a contribution of the internal qubit bath, $Q_0$, as well as a contribution of current (flux) fluctuations in the tank, $\lambda \hat{I}_T: Q = Q_0 + \lambda \hat{I}_T.$ An interaction with this bath, $H_{int} = - Q \sigma_z,$ has been integrated into the Hamiltonian (1).

We suppose that the frequency of the external field $\omega_0$  can be different from the energy splitting of the qubit $\omega_c = \sqrt{\Delta^2 + \varepsilon^2}$ with small detuning $\delta = \omega_0 - \omega_c, \delta \ll \omega_c.$ In the rotating frame of reference the qubit is described by the operators:
\begin{eqnarray}
X = {\Delta \over \omega_c}\sigma_x + {\varepsilon \over \omega_c}\sigma_z, \nonumber\\
Y = \sigma_y \cos\omega_0 t + \left({\Delta \over \omega_c}\sigma_z - {\varepsilon \over
\omega_c}\sigma_x\right)\sin\omega_0 t, \nonumber\\
Z =  \left({\Delta \over \omega_c}\sigma_z - {\varepsilon \over \omega_c}\sigma_x\right) \cos\omega_0 t -
\sigma_y \sin \omega_0 t
\end{eqnarray}
which have usual commutation rules: $[X,Y]_- = 2iZ,..$
In terms of these operators the Hamiltonian of the system can be rewritten as
\begin{equation}
H = {\omega_c \over 2} X - {A\over 2} Z -  \left[{\Delta \over \omega_c} \left(Z
\cos \omega_0t + Y \sin \omega_0t\right) + {\varepsilon \over \omega_c} X\right]Q + H_B.
\end{equation}
where $A$ is proportional to the amplitude of the driving force, $A = (\Delta F /\omega_c).$  Here we resort to the rotating wave approximation (RWA) and neglect fast oscillating terms. Taking into account an explicit time dependence of the operators $Y$ and $Z$ we derive the following Heisenberg equations $(\hbar = 1, k_B=1)$
\begin{eqnarray}
\dot{X} = A Y + 2{\Delta \over \omega_c}\left(Y \cos \omega_0t - Z \sin \omega_0t\right)(Q + f),
\nonumber\\
\dot{Y} = \delta Z - A  X - 2\left({\Delta \over \omega_c}X \cos \omega_0t -  {\varepsilon \over
\omega_c}Z\right)(Q + f), \nonumber\\
\dot{Z} = - \delta Y + 2\left({\Delta \over \omega_c}X \sin\omega_0t -  {\varepsilon \over \omega_c}Y\right)(Q +  f).
\end{eqnarray}
In the case of a Gaussian statistics of a free heat bath variables $Q^{(0)}$ or for a weak qubit-bath interaction a response of the heat bath on the action of the qubit is determined by the expression \cite{Efremov1}
\begin{equation}
Q(t) = Q^{(0)}(t) + \int dt_1 \varphi (t,t_1)\sigma_z(t_1),
\end{equation}
where
\begin{equation}
\sigma_z = {\Delta \over \omega_c} \left[
Z\cos\omega_0t + Y\sin \omega_0t\right] + {\varepsilon \over \omega_c} X.
\end{equation}
As in the Section II, a retarded Green function of free heat bath is denoted by $\varphi(t,t_1)$,
$
\varphi (t,t_1) = \langle i [Q^{(0)}(t),Q^{(0)}(t_1)]_-\rangle \theta (t-t_1)
$
with a respective Fourier transform (a susceptibility) $\chi(\omega )$.
This susceptibility, $\chi ( \omega ) =  \chi_0(\omega ) + \chi_T(\omega ),$
  incorporates a part  $\chi_0(\omega ),$ that is due to internal mechanisms of qubit decoherence, together with a resonant contribution of the tank, $\chi_T(\omega ),$
\begin{equation}
\chi_T(\omega ) = k^2 L_q I_q^2 {\omega_T^2 \over \omega_T^2 - \omega^2 - i \omega \gamma_T }.
\end{equation}
Besides that, the free heat bath is characterized by a correlation function $M(t,t_1),$
$
M(t,t_1) =\langle (1/2) [Q^{(0)}(t),Q^{(0)}(t_1)]_+\rangle,
$
and by a spectrum of equilibrium fluctuations $S(\omega)$ with temperature $T$,
\begin{equation}
S(\omega) = \int d\tau e^{i\omega \tau } M(\tau) = \chi^{\prime\prime}(\omega)
\coth \left({\omega \over 2T}\right).
\end{equation}
This spectrum $ S(\omega ) = S_0(\omega ) + S_T(\omega ),$  contains a part originated from the qubit interaction with its own heat bath, $S_0(\omega ),$
as well as a part, $S_T(\omega ),$ related to qubit coupling to the tank.
According to the fluctuation-dissipation theorem \cite{CallenWelton}, the equilibrium spectrum $S(\omega)$ is  proportional to the imaginary part of the heat bath susceptibility,
$\chi^{\prime\prime}(\omega).$

Following to the method outlined in the Appendix A we can rewrite the Heisenberg equations (23) in the form of quantum Langevin equations,
\begin{eqnarray}
\dot{X} = A Y + L_x + \xi_x + f_x,
\nonumber\\
\dot{Y} = \delta Z - A  X + L_y + \xi_y  + f_y, \nonumber\\
\dot{Z} = - \delta Y + L_z + \xi_z + f_z,
\end{eqnarray}
with the collision terms
\begin{eqnarray}
L_x(t) =  {2\Delta \over  \omega_c^2} \int dt_1 \{
\tilde{M}(t,t_1) i[Y(t)\cos\omega_0t - Z(t) \sin\omega_0t, \nonumber\\
\Delta Z(t_1)
\cos\omega_0t_1 +
\Delta Y(t_1)\sin\omega_0t_1 + \varepsilon X(t_1)]_- + \nonumber\\
\varphi(t,t_1) (1/2)[Y(t)\cos\omega_0t - Z(t) \sin\omega_0t, \nonumber\\
\Delta Z(t_1)
\cos\omega_0t_1 + \Delta Y(t_1)\sin\omega_0t_1 + \varepsilon X(t_1)]_+ \},\nonumber\\
L_y(t) = - {2 \over  \omega_c^2} \int dt_1 \{
\tilde{M}(t,t_1) i[\Delta X(t)\cos\omega_0t - \varepsilon Z(t), \nonumber\\
\Delta Z(t_1)
\cos\omega_0t_1 +
\Delta Y(t_1)\sin\omega_0t_1 + \varepsilon X(t_1)]_- + \nonumber\\
\varphi(t,t_1) (1/2)[\Delta X(t)\cos\omega_0t - \varepsilon Z(t), \nonumber\\
\Delta Z(t_1)
\cos\omega_0t_1 + \Delta Y(t_1)\sin\omega_0t_1 + \varepsilon X(t_1)]_+ \}, \nonumber\\
L_z(t) = {2 \over  \omega_c^2} \int dt_1 \{ \tilde{M}(t,t_1) i[\Delta
X(t)\sin\omega_0t - \varepsilon Y(t), \nonumber\\
\Delta Z(t_1) \cos\omega_0t_1 +
\Delta Y(t_1)\sin\omega_0t_1 + \varepsilon X(t_1)]_- + \nonumber\\
\varphi(t,t_1)\langle (1/2)[\Delta X(t)\sin\omega_0t - \varepsilon Y(t), \nonumber\\
\Delta Z(t_1)
\cos\omega_0t_1 + \Delta Y(t_1)\sin\omega_0t_1 + \varepsilon X(t_1)]_+ \},
\end{eqnarray}
 $\tilde{M}(\tau ) = M(\tau ) \theta(\tau ), \tau = t - t_1, $
and the fluctuation sources $\xi_x,\xi_y,\xi_z$. Definitions and correlation functions of these forces,  $ \xi_m(t) = \{ Q^{(0)}(t), {\cal {A}}_m {(t)} \}  (m = 1,2,3)  $ are presented in the Appendix A (see Eq.(A8)).
 Hereafter the digital indices $1,2,3$ correspond to the indices $x,y,z,$ respectively. The qubit operators ${\cal {A}}_m {(t)}$ are defined as follows (see also Eqs.(23)):
\begin{eqnarray}
{\cal {A}}_x {(t)} =  2{\Delta \over \omega_c} (Y \cos \omega_0t - Z \sin \omega_0t), \nonumber\\
{\cal {A}}_y {(t)}
 = - 2 \left({\Delta \over \omega_c}X \cos \omega_0t -  {\varepsilon \over
\omega_c}Z\right), \nonumber\\
{\cal {A}}_z {(t)} =  2 \left({\Delta \over \omega_c}X \sin\omega_0t -  {\varepsilon \over \omega_c}Y \right).
\end{eqnarray}
We also introduce the effective forces $f_x,f_y,f_z,:  f_m(t) =  {\cal {A}}_m {(t)}  f(t),$
which are necessary for calculating the linear response functions and susceptibilities of the qubit. After the calculations the auxiliary force $f(t)$ should be set equal to zero.

 The non-Markovian stochastic equations (28) can be simplified in the approximation of weak coupling between the qubit and the heat bath (Bloch-Redfield approximation).
With non-zero detuning, $ \delta \neq 0,$ the free evolution of the driven qubit (without coupling to a heat bath) is described by the equations $(\tau = t - t_1)$:
\begin{eqnarray}
X(t) = X(t_1) {\delta^2 + A^2 \cos\Omega_R \tau \over \Omega_R^2} + Y(t_1)
{A\over \Omega_R} \sin \Omega_R \tau + Z(t_1) A\delta {1 - \cos \Omega_R \tau \over \Omega_R^2}, \nonumber\\
Y(t) = Y(t_1) \cos \Omega_R \tau - X(t_1){A\over \Omega_R}\sin \Omega_R\tau + Z(t_1) {\delta \over \Omega_R} \sin \Omega_R \tau, \nonumber\\
Z(t) = Z(t_1) {A^2 + \delta^2 \cos\Omega_R \tau \over \Omega_R^2} - Y(t_1) {\delta \over \Omega_R} \sin \Omega_R \tau + X(t_1) A\delta {1 - \cos \Omega_R \tau \over \Omega_R^2},
\end{eqnarray}
where $\Omega_R$ is an effective Rabi frequency of the qubit,
\begin{equation}
\Omega_R = \sqrt{A^2 + \delta^2} = \sqrt{ \left({\Delta F\over \omega_c}\right)^2 + (\omega_0 - \omega_c)^2}.
\end{equation}
In the Bloch-Redfield approximation
  we can reduce the qubit operators taken at the moment $t$ to the operators at the moment $t_1$ using Eqs. (31), and, thereafter, calculate (anti)commutators of the simultaneous qubit operators using usual commutation rules: $[X(t_1),Y(t_1)]_- = 2i Z(t_1), [X(t_1),Y(t_1)]_+ = 0,..$ (see Appendix B). Neglecting fast oscillating terms in the collision integrals (29)  we derive the following equations for the qubit operators $X_1 = X, X_2 = Y, X_3 = Z$ in the rotating frame of reference $(m,n=1,2,3)$:
\begin{equation}
\dot {X}_m + \sum_n \Pi_{mn}X_n + \sum_n\int dt_1 \bar{\Gamma}_{mn}(t-t_1) X_n(t_1) =
\xi_m + f_m + \nu_m,
\end{equation}
where for nonzero elements of the matrix $\Pi$ we have: $\Pi_{12}= -\Pi_{21} = -A, \Pi_{23} = -\Pi_{32} = -\delta.$
The collision coefficients $\bar{\Gamma}_{mn}(\tau )$ are presented in the Appendix B.
For steady-state parameters $\nu_1= \nu_x,\nu_2=\nu_y,\nu_3 = \nu_z$  we obtain the following expressions:
\begin{eqnarray}
\nu_1= -  {\Delta^2 \over  \omega_c^2} \left[ {A^2 \over \Omega_R^2}
\chi^{\prime\prime}(\omega_0) + \right. \nonumber\\ \left.
{1\over 2} \left( 1 - {\delta \over \Omega_R}\right)^2
  \chi^{\prime\prime}(\omega_0 +\Omega_R) + {1\over 2} \left( 1 + {\delta \over \Omega_R}\right)^2 \chi^{\prime\prime}(\omega_0 -
\Omega_R) \right], \nonumber\\
\nu_2 =  {A \over \Omega_R}\left[ 2 {\delta \over \Omega_R}
{\varepsilon^2\over \omega_c^2} \left( \chi^{\prime}(0) - \chi^{\prime}(\Omega_R)\right) - {\delta \over \Omega_R}{\Delta^2 \over  \omega_c^2} \chi^{\prime}(\omega_0) +
\right. \nonumber\\ \left.
{\Delta^2 \over  2\omega_c^2} \left( 1 + {\delta \over \Omega_R}\right)\chi^{\prime}(\omega_0 -\Omega_R) - {\Delta^2 \over  2\omega_c^2} \left( 1 - {\delta \over \Omega_R}\right)\chi^{\prime}(\omega_0 +\Omega_R)\right], \nonumber\\
\nu_3 = {A \over \Omega_R}  \left[2 {\varepsilon^2\over \omega_c^2} \chi^{\prime\prime}(\Omega_R) + {\delta \over \Omega_R}{\Delta^2 \over  \omega_c^2} \chi^{\prime\prime}(\omega_0) + \right. \nonumber\\ \left.
{\Delta^2 \over  2\omega_c^2}
\left( 1 - {\delta \over \Omega_R}\right)^2
 \chi^{\prime\prime}(\omega_0 +\Omega_R) - {\Delta^2 \over  2\omega_c^2}
\left( 1 + {\delta \over \Omega_R}\right)^2 \chi^{\prime\prime}(\omega_0 -
\Omega_R) \right].
\end{eqnarray}
Here $\chi^{\prime}(\omega )$ and $ \chi^{\prime\prime}(\omega)$ are real and imaginary parts of the heat bath susceptibility  $\chi(\omega )$.

A formal solution of the equation (33) has the form
\begin{equation}
X_m(t) =  \sum_n \bar{G}_{mn}(t)X_n (0)  +   \sum_n \int dt_1 \bar{G}_{mn}(t-t_1)[\xi_n(t_1) + f_n(t_1) +\nu_n],
\end{equation}
where the last constant term describes the steady-state values of the average qubit
variables: $X_{1,0} = X_0 = (\delta /\Omega_R) P_0, X_{2,0} = Y_0 = 0, $ and $ X_{3,0} = Z_0 = (A/\Omega_R) P_0,$ with a polarization
\begin{equation}
P_0 = {4 (\varepsilon A/\Delta )^2 \chi^{\prime\prime}(\Omega_R) +
(\Omega_R - \delta)^2 \chi^{\prime\prime}(\omega_0 + \Omega_R) - (\Omega_R + \delta)^2 \chi^{\prime\prime}(\omega_0 - \Omega_R) \over
4 (\varepsilon A/\Delta )^2 S(\Omega_R) +
(\Omega_R - \delta)^2 S(\omega_0 + \Omega_R) + (\Omega_R + \delta)^2 S(\omega_0 - \Omega_R)}.
\end{equation}
For an exact resonance, $\omega_0 =\omega_c, \delta = 0,$
between the frequency of driving force $\omega_0$ and energy splitting of the qubit $\omega_c =
 \sqrt{\Delta^2 + \varepsilon^2}, $ the steady-state polarization $P_0$ is positive for the Ohmic or super-Ohmic heat bath, $\chi^{\prime\prime}(\omega) \sim \omega^r, r\geq 1. $  However, $P_0$ can be negative at non-zero detuning, $\delta \neq 0.$

Here we consider the case of strong driving when the Rabi frequency  $\Omega_R$ (32) is much more than qubit's relaxation rates.
Then, for Fourier transforms $G_{mn}(\omega)$ of the Green functions $\bar{G}_{mn}(\tau)$ incorporated into Eq.(35) we obtain:
\begin{eqnarray}
 G_{11}(\omega) = (\delta^2 - \omega^2)/D(\omega), G_{22}(\omega) = -\omega^2/D(\omega), G_{33}(\omega) = (A^2 - \omega^2)/D(\omega), \nonumber\\
 G_{12}(\omega) = -G_{21}(\omega) = -i\omega A/D(\omega),  G_{13}(\omega) = G_{31}(\omega) = \delta A/D(\omega), \nonumber\\
G_{23}(\omega) = -G_{32}(\omega) = -i\omega \delta /D(\omega),
\end{eqnarray}
 with a denominator
\begin{eqnarray}
D(\omega) =   i [\omega + i \Gamma_z(\omega)] [ \omega^2 - \Omega_R^2 + i \omega \Gamma(\omega )] \simeq  \nonumber\\
 i[\omega + i \Gamma_z(\omega)] [\omega - \Omega_R + i\Gamma(\omega)/2]
  [\omega + \Omega_R + i\Gamma(\omega)/2].
\end{eqnarray}
The coefficients $\Gamma_z(\omega), \Gamma(\omega)$, derived with Eqs.(B2),(B3) from the Appendix B,
play roles of frequency-dependent relaxation rates. These relaxation rates are even functions of $\omega, \Gamma_z(-\omega ) = \Gamma_z(\omega ), \Gamma(-\omega ) = \Gamma (\omega),$ and they are
 determined by the spectral density of the heat bath $S(\omega)$ (27):
\begin{eqnarray}
\Gamma_z(\omega) = {\varepsilon^2 \over \omega_c^2} {A^2 \over \Omega_R^2}
[ S(\omega + \Omega_R) + S (\omega - \Omega_R) ]  + \nonumber\\
{\Delta^2 \over 4 \omega_c^2}  \left( 1 - {\delta \over \Omega_R}\right)^2
 [ S(\omega + \omega_0 + \Omega_R) +
 S(\omega - \omega_0 - \Omega_R) ] + \nonumber\\
{\Delta^2 \over 4 \omega_c^2}  \left( 1 + {\delta \over \Omega_R}\right)^2
 [ S(\omega + \omega_0 - \Omega_R) +
 S(\omega - \omega_0 + \Omega_R) ];
\end{eqnarray}
\begin{eqnarray}
\Gamma(\omega ) = 2{\varepsilon^2 \over \omega_c^2} {A^2 \over \Omega_R^2}
S(\omega ) + \nonumber\\
2{\varepsilon^2 \over \omega_c^2} {\delta^2 \over \Omega_R^2}
\left[ \left(1 - {\Omega_R \over \omega } \right) S(\omega + \Omega_R ) +
\left(1 + {\Omega_R \over \omega } \right) S(\omega - \Omega_R )\right] + \nonumber\\
 {\Delta^2 \over 2 \omega_c^2} \left[ \left(1 - {\delta \over \Omega_R} \right)^2 S(\omega + \omega_0 ) + \left(1 + {\delta \over \Omega_R} \right)^2 S(\omega - \omega_0 ) \right] + \nonumber\\
{\Delta^2 \over 2 \omega_c^2}{A^2 \over \Omega_R^2} \left(1 - {\Omega_R \over \omega } \right) [S(\omega + \omega_0 + \Omega_R) +
 S(\omega - \omega_0 + \Omega_R) ] + \nonumber\\
{\Delta^2 \over 2 \omega_c^2}{A^2 \over \Omega_R^2} \left(1 + {\Omega_R \over \omega }\right) [S(\omega - \omega_0 - \Omega_R) +
 S(\omega + \omega_0 - \Omega_R) ].
\end{eqnarray}
We omit here frequency shifts of the qubit resulting from its interaction with the heat bath.
With Eqs.(37) we can calculate the retarded Green functions $\bar{G}_{mn}(\tau)$ defined at $\tau > 0$:
\begin{eqnarray}
\bar{G}_{xx}(\tau) = {\delta^2 \over \Omega_R^2} e^{-\Gamma_z \tau} + {A^2 \over \Omega_R^2} e^{-\Gamma \tau/2}\cos\Omega_R\tau, \nonumber\\
\bar{G}_{xy}(\tau) = -  \bar{G}_{yx}(\tau) = {A\over \Omega_R}  e^{-\Gamma \tau/2} \sin\Omega_R\tau,
\nonumber\\
\bar{G}_{xz}(\tau) = \bar{G}_{zx}(\tau) =  {\delta A\over \Omega_R^2} \left( e^{-\Gamma_z \tau}  -  e^{-\Gamma \tau /2}\cos\Omega_R\tau \right), \nonumber\\
\bar{G}_{yy}(\tau) = e^{-\Gamma \tau/ 2}\cos\Omega_R\tau, \nonumber\\
\bar{G}_{yz}(\tau) = - \bar{G}_{zy}(\tau) = {\delta \over \Omega_R} e^{-\Gamma \tau/2} \sin\Omega_R\tau, \nonumber\\
\bar{G}_{zz}(\tau) = {A^2 \over \Omega_R^2} e^{-\Gamma_z \tau } + {\delta^2 \over \Omega_R^2} e^{-\Gamma \tau/2 } \cos\Omega_R\tau ,
\end{eqnarray}
where $ \bar{G}_{mn}(0) = \delta_{mn}. $ Decay rates $\Gamma_z$ and $\Gamma$ are equal to  the functions $\Gamma_z(\omega)$ and $\Gamma(\omega)$  (39),(40) taken at zero frequency and at the Rabi frequency, respectively:
\begin{eqnarray}
\Gamma_z = \Gamma_z(0) =  2{\varepsilon^2 \over \omega_c^2} {A^2 \over \Omega_R^2}
S(\Omega_R) + \nonumber\\
{\Delta^2 \over 2 \omega_c^2} \left[ \left( 1 - {\delta \over \Omega_R}\right)^2
 S(\omega_0 + \Omega_R) + \left( 1 + {\delta \over \Omega_R}\right)^2
 S(\omega_0 - \Omega_R)\right];
\end{eqnarray}
\begin{eqnarray}
\Gamma = \Gamma(\Omega_R) = \Gamma_z + 4{\varepsilon^2 \over \omega_c^2} {\delta^2 \over \Omega_R^2} S(0) +
2{\Delta^2 \over \omega_c^2} {A^2 \over \Omega_R^2}
S(\omega_0 ).
\end{eqnarray}
The decay rate $\Gamma/2$ is related to the rate $T_1^{-1}$ from
Ref. \cite{Smirnov1}, a notation for $\Gamma_z$ remains the same.
It should be noted also that the heat bath operator, $Q(t),$
defined in the present paper  differs from the same operator $Q$
from the paper \cite{Smirnov1}, by the factor $1/2$ (see also
\cite{footnote1}). Because of this, to compare our results with
results of the above-mentioned article we have to divide our
spectral function of the heat bath $S(\omega)$ and our
susceptibility $\chi (\omega )$ by four to get the spectral
function and the susceptibility of the heat bath used in Ref
\cite{Smirnov1}.

\section{ Dissipative dynamics and a linear response of the qubit.}

It follows from Eq. (35) that an  evolution of the averaged qubit operator $\langle X_m(t)\rangle , m = 1,2,3, $
from its initial condition $ \langle X_m(0)\rangle $ to the steady-state value $ X_{m,0} $ is governed by the corresponding Green functions $ \bar{G}_{mn}(t) $ (41):
\begin{equation}
\langle X_m(t)\rangle =  X_{m,0}  + \sum_n \bar{G}_{mn}(t) [ \langle X_n(0) \rangle  - X_{n,0} ].
\end{equation}
In particular, if the qubit starts from the ground state $|0\rangle $ of the Hamiltonian $H_0 = (\Delta / 2) \sigma_x + (\varepsilon / 2) \sigma_z = (\omega_c/2) X, $ where $  \langle 0 | X(0)| 0 \rangle = -1,
\langle 0 | Y(0)| 0 \rangle =  \langle 0 | Z(0)| 0 \rangle =  0,$ a probability to find the qubit in the excited state,  $ P_E = ( 1 + \langle X \rangle )/2, $
\begin{eqnarray}
P_E(t) = {1\over 2} \left( 1 + {\delta \over \Omega_R} P_0 \right) \left( 1 - {\delta^2 \over \Omega_R^2} e^{-\Gamma_z t} - {A^2 \over \Omega_R^2} e^{-\Gamma t/2}\cos\Omega_Rt \right) - \nonumber\\
{\delta A^2\over 2\Omega_R^3} \left( e^{-\Gamma_z t}  -  e^{-\Gamma t /2} \cos\Omega_Rt \right),
 \end{eqnarray}
oscillates with the Rabi frequency $\Omega_R $ (32) and relaxes to the steady-state value
$P_{E, st} = (1/2)[ 1 + (\delta /\Omega_R)P_0].$ In the case of resonant driving when $\omega_0 = \omega_c, \delta = 0, $ the energy levels of the qubit  are populated equally in the steady state, $P_{E, st} = 1/2.$
At the same initial conditions the averaged current in the qubit loop,
$ \langle \hat{I}_q(t) \rangle = I_q \langle \sigma_z(t) \rangle ,$
oscillates not only with the Rabi frequency $\Omega_R$, but also with the frequency of driving force $\omega_0,$ as well as with frequencies $\omega_0 \pm \Omega_R:$
\begin{eqnarray}
\langle \hat{I}_q(t) \rangle = {\Delta \over \omega_c}{A\over \Omega_R} I_q \cos\omega_0 t \left[
P_0 \left( 1 - {A^2 \over \Omega_R^2} e^{-\Gamma_z t } - {\delta^2 \over \Omega_R^2} e^{-\Gamma t/2 } \cos\Omega_R t  \right) - \right. \nonumber\\ \left.
- {\delta \over \Omega_R} \left( 1 + {\delta \over \Omega_R} P_0 \right) \left( e^{-\Gamma_z t}  -  e^{-\Gamma t /2}\cos\Omega_Rt \right)\right] + \nonumber\\
{\Delta \over \omega_c}{A\over \Omega_R} I_q \sin\omega_0 t  \sin\Omega_Rt  e^{-\Gamma t/2 }  - \nonumber\\
{\varepsilon \over \omega_c} I_q  \left( 1 + {\delta \over \Omega_R} P_0 \right)
\left( {\delta^2 \over \Omega_R^2} e^{-\Gamma_z t} + {A^2 \over \Omega_R^2} e^{-\Gamma t/2}\cos\Omega_Rt \right) -   \nonumber\\
{\varepsilon \over \omega_c} I_q {\delta \over \Omega_R} P_0 \left[ {A^2 \over \Omega_R^2} \left( e^{-\Gamma_z t}  -  e^{-\Gamma t /2}\cos\Omega_Rt \right) -1 \right].
\end{eqnarray}
The qubit starts with an initial current, $\langle \hat{I}_q(0) \rangle = - (\varepsilon/\omega_c)I_q, $ corresponding to the ground state with a non-zero bias $\varepsilon,$ and ends  at $ t \gg \Gamma_z^{-1}, 2/\Gamma$   with the  steady-state current oscillating with the frequency of the driving force $\omega_0:$
\begin{equation}
\langle \hat{I}_q(t) \rangle_{st} = I_q P_0  \left( {\Delta \over \omega_c}{A\over \Omega_R} \cos\omega_0 t +
{\varepsilon \over \omega_c} {\delta \over \Omega_R}\right) .
\end{equation}
Interestingly, there are no signs of the Rabi frequency in oscillations of the steady-state qubit current.
We emphasize that  a relaxation of the population difference $P_E$ and a decay of the qubit current $ \langle \hat{I}_q(t) \rangle $ are determined by both damping rates $\Gamma_z$ and $\Gamma /2$ (42),(43). It is worth to note also, that zero-frequency fluctuations of the heat bath, such as 1/f-noise, which are described by the spectral function $S(0),$ contribute to the decay rate $\Gamma $ (43) in the case of non-zero detuning $\delta$ and non-zero bias $\varepsilon. $

Equations (35) averaged over the initial state of the qubit and over the thermodynamically-equilibrium initial state of the heat bath allow us to calculate a linear response of the qubit on the action of small external force $f(t),$ namely, a functional derivative $\langle \delta \sigma_z(t) / \delta f(t')\rangle$ or its Fourier transform, a magnetic susceptibility $\chi_{zz}(\omega)$ (10).
With Eq.(25) we obtain the following expression:
\begin{equation}
\langle {\delta \sigma_z(t) \over \delta f(t')}\rangle = {\Delta \over \omega_c}\langle {\delta Z(t) \over \delta f(t')} \rangle \cos\omega_0t + {\Delta \over \omega_c}\langle {\delta Y(t) \over \delta f(t')} \rangle \sin\omega_0t +
{\varepsilon \over \omega_c}\langle {\delta X(t) \over \delta f(t')} \rangle
\end{equation}
The derivatives of the qubit operators in the rotating frame of reference, $ \langle \delta X_m(t)/\delta f(t') \rangle$, can be found from the averaged equation (35) taking into account formulas  for the forces $f_n(t): f_n(t) =  {\cal {A}}_n {(t)}  f(t),$ where $ {\cal {A}}_n {(t)} $ are defined by Eqs. (30).
The mean values of the averaged qubit variables $\langle X \rangle,
\langle Y \rangle,\langle Z \rangle $ in Eqs.(30) should be replaced in the process by their steady-state values $X_0 = (\delta /\Omega_R) P_0, Y_0 = 0, $ and $ Z_0 = (A/\Omega_R) P_0,$ where  the polarization $P_0$ is given by Eq. (36).  Then, for the magnetic susceptibility of the qubit we obtain the following result:
\begin{eqnarray}
\chi_{zz}(\omega ) = -  {\varepsilon^2 \over \omega_c^2}  {A^2 \over \Omega_R^2} P_0
\left( {1\over \omega - \Omega_R + i \Gamma/2} - {1\over \omega + \Omega_R + i \Gamma/2}\right) - \nonumber\\
{\Delta^2 \over 4 \omega_c^2} P_0 \left[1 + 2 {\delta (\omega + \omega_0)\over \Omega_R^2} + {\delta^2 \over \Omega_R^2}\right]
\left( {1\over \omega +\omega_0 - \Omega_R + i \Gamma/2} - {1\over \omega + \omega_0 +
\Omega_R + i \Gamma/2}\right) - \nonumber\\
{\Delta^2 \over 4 \omega_c^2} P_0 \left[1 - 2 {\delta (\omega - \omega_0)\over \Omega_R^2} + {\delta^2 \over \Omega_R^2}\right]
\left( {1\over \omega - \omega_0 - \Omega_R + i \Gamma/2} - {1\over \omega - \omega_0 +
\Omega_R + i \Gamma/2}\right).
\end{eqnarray}
The imaginary part of this susceptibility which defines absorption properties of the driven qubit peaks at the Rabi frequency $\Omega_R$ as well as at frequencies $\omega_0\pm \Omega_R:$
\begin{eqnarray}
\chi_{zz}^{\prime\prime}(\omega )  =  {\varepsilon^2 \over \omega_c^2}  {A^2 \over \Omega_R^2} P_0
\left[ {\Gamma/2\over (\omega - \Omega_R)^2 +  (\Gamma/2)^2} - {\Gamma/2\over (\omega + \Omega_R)^2 +  (\Gamma/2)^2}\right] + \nonumber\\
{\Delta^2 \over 4 \omega_c^2} P_0 \left( 1 + {\delta \over \Omega_R}\right)^2
\left[ {\Gamma/2\over (\omega + \omega_0 - \Omega_R)^2 +  (\Gamma/2)^2} - {\Gamma/2\over (\omega -\omega_0 + \Omega_R)^2 +  (\Gamma/2)^2}\right] + \nonumber\\
{\Delta^2 \over 4 \omega_c^2} P_0 \left( 1 - {\delta \over \Omega_R}\right)^2
\left[ {\Gamma/2\over (\omega - \omega_0 - \Omega_R)^2 +  (\Gamma/2)^2} - {\Gamma/2\over (\omega +\omega_0 + \Omega_R)^2 +  (\Gamma/2)^2}\right].
\end{eqnarray}
An absorption of weak signal energy by the qubit is determined by the function $U(\omega ) = \omega \chi_{zz}^{\prime\prime}(\omega )$ \cite{Fain1}. This is evident from the formula (50) that $U(\omega )$ can be negative at the positive frequency $\omega_0 - \Omega_R.$ It means that a weak signal having this frequency will be amplified by the strongly driven qubit.
The low-frequency part of the qubit susceptibility which affects the resonant frequency of the tank (see Eq.(19)) has the form
\begin{equation}
\chi_{zz}^{\prime}(\omega ) = - {\varepsilon^2 \over \omega_c^2}  {A^2 \over \Omega_R^2} P_0 \left[ {\omega - \Omega_R \over (\omega - \Omega_R)^2 +  (\Gamma/2)^2} - {\omega+ \Omega_R\over (\omega + \Omega_R)^2 +  (\Gamma/2)^2}\right].
\end{equation}
The angle $\Theta $ between a voltage and a current in the tank coupled to the qubit  (14) is  determined by the susceptibility of the qubit (50),(51) taken at the tank frequency $\omega_T:$
\begin{equation}
\tan \Theta = - k^2 L_qI_q^2 {\varepsilon^2 \over \omega_c^2}  {A^2 \over \Omega_R^2} P_0
{ \Omega_R - \omega_T  \over (\Omega_R - \omega_T )^2 +  (\Gamma/2)^2} \times
{ \omega_T \over \bar{\gamma}_T}.
\end{equation}
Here $\bar{\gamma}_T$,
\begin{equation}
\bar{\gamma}_T = \gamma_T + k^2 L_qI_q^2 {\varepsilon^2 \over \omega_c^2}  {A^2 \over \Omega_R^2} P_0 \omega_T {\Gamma/2\over (\Omega_R - \omega_T )^2 +  (\Gamma/2)^2}
\end{equation}
is an effective damping rate of the tank in the presence of the qubit. \\

\section{Nonequilibrium spectra of the qubit and the tank.}

Here we will calculate the total nonequilibrium spectrum of the qubit fluctuations, $S_{zz}(\omega )$  (17), for the case of exact resonance between energy  splitting of the qubit, $\omega_c = \sqrt{\Delta^2 + \varepsilon^2},$ and the frequency $\omega_0$ of the driving field: $\delta = \omega_0 - \omega_c = 0.$
Calculations of the spectrum for non-zero detuning are straightforward, but cumbersome enough.
The part of the qubit spectrum $S_{zz,0}(\omega ),$  which results from internal decoherence mechanisms of the qubit (coupling to  the bath  $Q_0$), can be easily found from the expression for the total spectrum $S_{zz}.$ To do that we have to replace the total spectrum of the bath, $S(\omega)$ by the spectrum of  the internal bath $S_0(\omega)$ in the expressions  (C2)-(C7) for the spectra of fluctuation forces given in the Appendix C. It should be emphasized that damping rates of the qubit are determined nevertheless by the total spectrum $S(\omega)$ (27) of the dissipative environment.

In view of Eq.(25) the correlator of $\sigma_z$-operators of the qubit averaged over fast oscillations
can be expressed in terms of qubit's correlation functions in the rotating frame $(\tau = t-t'):$
\begin{eqnarray}
\langle {1\over 2} \left[\sigma_z(t), \sigma_z(t')\right]_+\rangle =  {\varepsilon^2\over \omega_c^2} \langle {1\over 2} \left[X(t), X(t')\right]_+\rangle + \nonumber\\
{\Delta^2\over 2\omega_c^2}\left\{ \langle {1\over 2} \left[Z(t), Z(t')\right]_+\rangle
\cos\omega_0\tau + \langle {1\over 2} \left[Y(t), Y(t')\right]_+\rangle
\cos\omega_0\tau + \right. \nonumber\\ \left.
\langle {1\over 2} \left[Y(t), Z(t')\right]_+\rangle
\sin\omega_0\tau - \langle {1\over 2} \left[Z(t), Y(t')\right]_+\rangle
\sin\omega_0\tau \right\}.
\end{eqnarray}
Correlators of the qubit variables in the rotating frame are determined by the correlation functions of the fluctuation forces (see  a stochastic part of Eq.(35))
\begin{equation}
\langle {1\over 2} \left[X_m(t), X_k(t')\right]_+\rangle = \int dt_1 dt_2
\sum_{nl}
\bar{G}_{mn}(t,t_1)\bar{G}_{kl}(t,t_2) \langle {1\over 2} \left[\xi_n(t_1), \xi_l(t_2)\right]_+\rangle,
\end{equation}
where the Green functions, $\bar{G}_{mn}(\tau),$ are defined by Eqs.(41).
Fourier transforms of the qubit correlation functions, $\Lambda_{mk}(\omega)   $, and the correlator of the fluctuation forces, $ K_{nl}(\omega )$,
$$ \langle {1\over 2} \left[X_m(t), X_k(t')\right]_+\rangle = \int {d\omega \over 2 \pi} e^{-i\omega (t-t')}\Lambda_{mk}(\omega); $$ $$
\langle {1\over 2} \left[\xi_n(t), \xi_l(t')\right]_+\rangle =
\int {d\omega \over 2 \pi} e^{-i\omega (t-t')}K_{nl}(\omega ),$$
are related according to the equation
\begin{equation}
\Lambda_{mk}(\omega) =  \sum_{lq} G_{mn}(\omega )G_{kl}(-\omega ) K_{nl}(\omega )
\end{equation}
with $G_{mn}(\omega )$ from Eqs.(37).
Correlators of fluctuation forces, $\langle (1/2) \left[\xi_n(t), \xi_l(t')\right]_+\rangle $, are calculated according to the procedure given in the Appendix A (see Eq.(A11)). The expressions for spectral functions of the fluctuation forces, $ K_{nl}(\omega ),$
are presented in the Appendix C.

For the spectrum of qubit fluctuations $S_{zz}(\omega )$ we find from Eqs.(17),(54):
\begin{eqnarray}
S_{zz}(\omega ) =  {\varepsilon^2\over \omega_c^2} \Lambda_{xx}(\omega) + \nonumber\\
{\Delta^2 \over 4 \omega_c^2}
[ \Lambda_{zz}(\omega + \omega_0) + \Lambda_{yy}(\omega + \omega_0) - i \Lambda_{yz}(\omega + \omega_0) +  i \Lambda_{zy}(\omega + \omega_0)] + \nonumber\\
{\Delta^2 \over 4 \omega_c^2}
[ \Lambda_{zz}(\omega - \omega_0) + \Lambda_{yy}(\omega - \omega_0) + i \Lambda_{yz}(\omega - \omega_0) -  i \Lambda_{zy}(\omega + \omega_0)].
\end{eqnarray}
It follows from Eqs.(57),(37), that the spectral functions of the qubit operators in the rotating frame, $ \Lambda_{mk}(\omega),  (m,k = x,y,z)  $ are determined by the spectra of fluctuation forces  $K_{nl}(\omega )$:
\begin{eqnarray}
\Lambda_{xx}(\omega) = \omega^2 [ \omega^2 K_{xx}(\omega )  + \Omega_R^2 K_{yy}(\omega ) - 2i\omega A  K_{xy}(\omega )]/|D(\omega)|^2, \nonumber\\
\Lambda_{yy}(\omega) = \omega^2 [ \Omega_R^2 K_{xx}(\omega )  + \omega^2 K_{yy}(\omega ) - 2i\omega A  K_{xy}(\omega )] /|D(\omega)|^2, \nonumber\\
\Lambda_{zz}(\omega) = (\omega^2 - \Omega_R^2)^2 K_{zz}(\omega ) /|D(\omega)|^2, \nonumber\\
\Lambda_{yz}(\omega) = i \omega (\Omega_R^2 - \omega^2) [  A K_{xz}(\omega ) + i \omega K_{yz}(\omega )] / |D(\omega)|^2,
\end{eqnarray}
with $\Lambda_{zy}(\omega)=- \Lambda_{yz}(\omega).$ Here
\begin{equation}
|D(\omega)|^2 = [\omega^2 + \Gamma_z^2(\omega )] [ (\omega^2 - \Omega_R^2)^2 + \omega^2 \Gamma^2(\omega )]
\end{equation}
is the modulus square of the Green function denominator (38).
Combining Eqs.(57)-(59) with the formulas (C2)-(C7) from the Appendix C we obtain the nonequilibrium spectrum of qubit fluctuations $S_{zz}(\omega ):$
\begin{eqnarray}
S_{zz}(\omega ) = {\varepsilon^2\over 2\omega_c^2} { W_R (\omega ) \over
 (\omega^2 - \Omega_R^2)^2 + \omega^2 \Gamma^2(\omega ) }
 + \nonumber\\
{\Delta^2 \over  4\omega_c^2}  { W(\omega_0 + \omega ) \over
 (\omega + \omega_0)^2 + \Gamma_z^2(\omega + \omega_0 ) } \times
{1\over   [(\omega + \omega_0)^2  - \Omega_R^2]^2 + \omega^2 \Gamma^2(\omega + \omega_0) }
 + \nonumber\\
{\Delta^2 \over  4\omega_c^2}  { W(\omega_0 - \omega ) \over
(\omega - \omega_0)^2 + \Gamma_z^2(\omega - \omega_0 ) } \times
{1\over  [ (\omega - \omega_0)^2  - \Omega_R^2]^2 + \omega^2 \Gamma^2(\omega - \omega_0) }
\end{eqnarray}
Here the frequency-dependent damping rates $\Gamma_z(\omega), \Gamma(\omega)$ are defined by Eqs. (39),(40).  Functions $W_R(\omega)$ and $W(\omega )$ are given by the expressions
\begin{eqnarray}
W_R(\omega ) = 8  \Omega_R^2 {\varepsilon^2\over \omega_c^2} S(\omega ) +
 2 \omega^2 {\Delta^2\over \omega_c^2} [ S(\omega + \omega_0) + S(\omega - \omega_0)] +
\nonumber\\
{\Delta^2\over \omega_c^2} (\omega - \Omega_R)^2 [  S(\omega + \omega_0+ \Omega_R) + S(\omega - \omega_0 + \Omega_R) - \nonumber\\
P_0 \chi^{\prime\prime}(\omega + \omega_0 + \Omega_R) - P_0 \chi^{\prime\prime}(\omega - \omega_0 + \Omega_R)] + \nonumber\\
{\Delta^2\over \omega_c^2} (\omega + \Omega_R)^2 [  S(\omega - \omega_0 - \Omega_R) + S(\omega + \omega_0 - \Omega_R) + \nonumber\\
P_0 \chi^{\prime\prime}(\omega - \omega_0 - \Omega_R) + P_0 \chi^{\prime\prime}(\omega + \omega_0 - \Omega_R)],
\end{eqnarray}
\begin{eqnarray}
W(\omega ) = 4 \omega^4  {\varepsilon^2\over \omega_c^2} S(\omega ) +
\omega^2 \Omega_R^2  {\Delta^2\over \omega_c^2} [ S(\omega + \omega_0) + S(\omega - \omega_0)] + \nonumber\\
2 {\varepsilon^2\over \omega_c^2} (\omega^2 - \Omega_R^2)^2 [ S(\omega + \Omega_R) -
  P_0 \chi^{\prime\prime}(\omega + \Omega_R) + S(\omega - \Omega_R) +
  P_0 \chi^{\prime\prime}(\omega - \Omega_R)] + \nonumber\\
\Omega_R^2  {\Delta^2\over 2\omega_c^2}  (\omega - \Omega_R)^2 [
S(\omega + \omega_0 + \Omega_R) -   P_0 \chi^{\prime\prime}(\omega + \omega_0 + \Omega_R) ] + \nonumber\\
\Omega_R^2  {\Delta^2\over 2\omega_c^2}  (\omega + \Omega_R)^2 [ S(\omega + \omega_0 - \Omega_R) +   P_0 \chi^{\prime\prime}(\omega + \omega_0 - \Omega_R) ] + \nonumber\\
{\Delta^2\over 2\omega_c^2}  (\omega - \Omega_R)^2 (2\omega + \Omega_R)^2
[ S(\omega - \omega_0 + \Omega_R) -   P_0 \chi^{\prime\prime}(\omega - \omega_0 + \Omega_R) ] + \nonumber\\
{\Delta^2\over 2\omega_c^2}  (\omega + \Omega_R)^2 (2\omega - \Omega_R)^2
[ S(\omega - \omega_0 - \Omega_R) +   P_0 \chi^{\prime\prime}(\omega - \omega_0 - \Omega_R) ].
\end{eqnarray}
It follows herefrom that the qubit spectrum is an even function of $ \omega: S_{zz}(-\omega ) =
S_{zz}(\omega ). $
In the regime of strong driving, when $\Omega_R \gg \Gamma,\Gamma_z,$ the qubit spectrum $S_{zz}(\omega)$ taken at positive frequencies consists of three Lorentzian peaks centered at
frequencies $\Omega_R, \omega_0 - \Omega_R,$ and $\omega_0 + \Omega_R$ with the linewidth $\Gamma/2$ each, with $\Gamma = \Gamma(\Omega_R) $ (43), complemented by the additional peak located exactly at the frequency of the driving force $\omega_0$. In the present section $\omega_0$ is equal to the frequency of quantum beatings $\omega_c$. The additional peak has a different linewidth, $\Gamma_z = \Gamma_z(0)$ (42), and it is absent in the absorption spectrum determined by the function $\chi_{zz}^{\prime\prime}(\omega )$ (50). The current in the qubit loop is described by the operator
$ \hat{I}_q = I_q \sigma_z$. Because of this the spectrum of current fluctuations of the qubit, $S_I(\omega),$  is proportional to the spectrum $S_{zz}(\omega ): S_{I}(\omega ) = \langle (1/2)[\hat{I}_q(\omega ), \hat{I}_q ]_+ \rangle = I_q^2 S_{zz}(\omega ), $ where $I_q$  is the value of the persistent current in the qubit loop.

In view of the facts, that $\omega_0 \gg \Omega_R,$ and $ W_R(\Omega_R) \simeq 4 \Gamma\Omega_R^2 , W(0) \simeq
2\Gamma_z \Omega_R^4, W(\Omega_R) \simeq 2\Gamma \Omega_R^4, $
 and considering the heat bath with temperature that is greater than the energy of a Rabi quantum, $T \gg \hbar \Omega_R,$ we find the simple formula for the nonequilibrium spectrum of qubit fluctuations:
\begin{eqnarray}
S_{zz}(\omega ) = 2 {\varepsilon^2 \over \omega_c^2} { \Omega_R^2 \Gamma \over
(\omega^2 - \Omega_R^2)^2 + \omega^2 \Gamma^2 }  + \nonumber\\
{\Delta^2\over 2\omega_c^2}
{ \Gamma_z \over
(\omega + \omega_0)^2 + \Gamma_z^2 } + {\Delta^2\over 2\omega_c^2}  { \Gamma_z \over (\omega - \omega_0)^2 + \Gamma_z^2 } + \nonumber\\
{\Delta^2\over 2\omega_c^2}  {\Gamma \over
[ (\omega + \omega_0)^2  - \Omega_R^2]^2 + \omega^2 \Gamma^2 } +
{\Delta^2\over 2\omega_c^2}  {\Gamma \over
[ (\omega - \omega_0)^2  - \Omega_R^2]^2 + \omega^2 \Gamma^2 },
\end{eqnarray}
where the decay rates can be found from Eqs.(42),(43) at zero detuning $\delta = 0$:
\begin{eqnarray}
\Gamma_z = 2{\varepsilon^2 \over \omega_c^2} S(\Omega_R) +{\Delta^2 \over  \omega_c^2} S(\omega_0 ), \nonumber\\
\Gamma = 2{\varepsilon^2 \over \omega_c^2} S(\Omega_R) +3{\Delta^2 \over  \omega_c^2} S(\omega_0 ).
\end{eqnarray}
As for the driven atom \cite{Mollow1} the spectrum of our two-level system is double peaked at the frequencies
$\omega_0 \pm \Omega_R,$ but, besides that, we have a peak at the Rabi frequency with the intensity that is proportional to the bias squared $\varepsilon^2.$
The low-frequency part of the spectrum $S_{zz}(\omega )$ gives a significant contribution to the voltage spectrum of the tank $S_V(\omega ) $ (18). We recall also that the internal heat bath only should be taken into account in the process of calculating the spectra of fluctuation forces
$K_{nl}(\omega )$ (C2)-(C7). It means that in expressions  (61),(62) for $W_R(\omega ), W(\omega ) $  we have to
extract fluctuations of the tank from the total heat bath spectrum, $S(\omega ),$ and
substitute this spectrum  for the spectrum $S_0(\omega )$ related to the internal bath.
The damping rates $\Gamma$ and $\Gamma_z$ in numerators of Eq. (63) are originated exactly from the function $W(\omega )$ (62). For calculating the spectrum $S_{zz,0}(\omega ) $
 it is necessary to replace these rates by the coefficients depending on the spectrum of the internal bath only, namely, $\Gamma_{z,0}$ and $\Gamma_0,$ where, for example, $ \Gamma_0 =
2(\varepsilon^2 / \omega_c^2) S_0(\Omega_R) +3(\Delta^2 / \omega_c^2) S_0(\omega_0 ). $
The decay rates in the denominators of the spectrum $S_{zz,0}(\omega)$ remain the same, because both the internal mechanisms and the tank fluctuations contribute to the linewidth of the qubit.
As a result, for the low-frequency part of the qubit spectrum $S_{zz,0}(\omega ) $ we find
\begin{equation}
S_{zz,0}(\omega ) = 2{\varepsilon^2 \over \omega_c^2} { \Omega_R^2 \Gamma_0 \over
(\omega^2 - \Omega_R^2)^2 + \omega^2 \Gamma^2 }
\end{equation}
It should be noted that all of these nuances with replacing $S(\omega ) $ to $S_0(\omega ) $ and $\Gamma $ to $\Gamma_0$  in the spectrum $S_{zz}(\omega )$ (63) are important only when the contribution of the tank into the fluctuations and decoherence of the qubit is quite significant. For weak qubit-tank inductive coupling a strong  influence of the tank on the qubit coherence can take place only near the exact resonance between the Rabi frequency $\Omega_R$ and the resonance frequency of the tank $\omega_T$. It is necessary to develop more rigorous theory to study a close proximity of this point.

The total spectrum of voltage fluctuations in the tank, $S_V(\omega )$ (18), incorporates a contribution of the tank noise, $ S_{VT}(\omega ), $ together with a low-frequency contribution of the driven qubit, $ S_{VQ}(\omega): S_V(\omega ) = S_{VT}(\omega ) + S_{VQ}(\omega).$
At temperatures $T \gg \omega_T $ the tank contribution to the voltage spectrum  is proportional to the spectral function $S_b(\omega) = T \gamma_T  /\omega_T $ (7):
\begin{equation}
S_{VT}(\omega ) =  2 {\omega^2 \over C_T}  { T \gamma_T
 \over
(  \bar{\omega}_T^2  -  \omega^2 )^2 + \omega^2  \bar{\gamma}_T^2 },
\end{equation}
where $\bar{\omega }_T$ is the resonance frequency of the tank (19) shifted due to qubit-tank coupling,
\begin{equation}
\bar{\omega }_T = \omega_T\sqrt{ 1 + k^2L_q I_q^2
{\varepsilon^2 \over \omega_c^2}  P_0  {\omega_T - \Omega_R \over (\omega_T - \Omega_R)^2 +  (\Gamma/2)^2} }.
\end{equation}
We use here Eq.(51) for the function $\chi^{\prime}(\omega_T)$.The linewidth of the tank, $\bar{\gamma}_T$, modified by the qubit, is defined therewith by Eq.(53).

For the qubit part we find from Eqs. (18),(65):
\begin{equation}
S_{VQ}(\omega ) = 2 {\varepsilon^2 \over \omega_c^2} k^2 {L_q I_q^2  \over C_T} \omega^2 \Gamma_0  {  \omega_T^2 \over
(\bar{\omega}_T^2  -  \omega^2)^2 + \omega^2  \bar{\gamma}^2_T
} \times
 { \Omega_R^2  \over
(\omega^2 - \Omega_R^2)^2 + \omega^2 \Gamma^2 }  .
\end{equation}
Measurements of the voltage fluctuations are performed within the linewidth of the tank: $\omega \simeq \bar{\omega}_T \pm \gamma_T.$ In this frequency range a signal-to-noise ratio demonstrates a resonant behaviour as a function of the Rabi frequency $\Omega_R:$
\begin{equation}
{ S_{VQ}(\omega )\over S_{VT}(\omega )}_{|\omega = \omega_T} =  {\varepsilon^2 \over \omega_c^2} k^2 {L_q I_q^2 \over T} {\Gamma_0 \over \gamma_T} { \omega_T^2 \Omega_R^2  \over
(\omega_T^2 - \Omega_R^2)^2 + \omega_T^2 \Gamma^2 }.
\end{equation}
In this expression  we have a ratio of bare damping rates of the qubit, $\Gamma_0,$ and the tank,$\gamma_T.$
Besides the part $\Gamma_0$, related to the contribution of the internal heat bath into qubit decoherence, the  total decay rate of the qubit $\Gamma$ (64) contains also a term, $\Gamma_T$, which describes a tank share in the qubit damping: $\Gamma = \Gamma_0 + \Gamma_T,$ with
\begin{equation}
\Gamma_T = 4 k^2 L_q I_q^2 {\varepsilon^2 \over \omega_c^2} \omega_T^2
 { T \gamma_T \over
(\omega_T^2 - \Omega_R^2)^2 + \Omega_R^2 \gamma_T^2 }.
\end{equation}
This rate reflects a backaction of the detector (LC circuit) on the quantum bit that accompanies an acquisition of any information from the qubit. Both parameters, $ S_{VQ} / S_{VT}$ and $\Gamma_T$ reach the maximums when the Rabi frequency $\Omega_R$ is about the resonant frequency of the tank $\omega_T.$
However, the signal-to-noise ratio (69) as a function of the Rabi frequency $\Omega_R $ has a linewidth  that is determined by the damping rate of the qubit $\Gamma$, whereas the tank contribution to the qubit decoherence (70) is localized in the narrower range of Rabi frequencies which is of order of the tank linewidth $\gamma_T, \gamma_T \ll \Gamma.$  Measurements performed at the Rabi frequencies that are out of this range, $\Omega_R > \omega_T + \gamma_T,$ or $ \Omega_R < \omega_T - \gamma_T,$  can demonstrate a good efficiency, $ S_{VQ}/S_{VT} > 1$,  without introducing strong decoherence in the qubit. To observe a signature of Rabi oscillations in the spectrum of voltage fluctuations $S_V(\omega)$ the decay rate $\Gamma$ and, all the more, the measurement-induced rate $\Gamma_T$ should be appreciably less than the Rabi frequency $\Omega_R \simeq \omega_T:
 \Gamma_T / \omega_T \ll 1.$ For the flux qubit measured in Ref.\cite{Il'ichev1} we have the following set of parameters: $ L_q = 24 pH, I_q \simeq 500 nA, Q_T = \omega_T/2\gamma_T = 1850, T = 10 mK, \omega_T/2\pi = 6.284 MHz, \hbar \omega_T = 4.16 \times 10^{-27} J, $ so that $ T/\hbar \omega_T = 33, L_q I_q^2 = 6 \times 10^{-24} J, L_qI_q^2/\hbar \omega_T = 1440, \omega_T/\gamma_T = 3700.$ If we take a  value of the coupling parameter squared,  $k^2 \sim  10^{-3},$  from Ref.  \cite{Il'ichev1} and suppose that $\varepsilon/\Delta \simeq 1/126,$ then for the measurement-induced damping rate we obtain the ratio: $\Gamma_T /\omega_T \simeq 0.8 \times 10^{-2},$ at the point where $\Omega_R - \omega_T \simeq \Gamma/2 \simeq 0.01 \omega_T.$ We use here the decay rate, $\Gamma = 0.02 \omega_T \ll \omega_T$, measured in Ref. \cite{Il'ichev1}.
At these conditions  the signal-to-noise ratio (69)  is of order $0.5$, and the detector-induced decoherence of the qubit,
$\Gamma_T,$  as well as the total rate $\Gamma$ are much less than the Rabi frequency of the qubit $\Omega_R.$
It means that the spectroscopic observation of Rabi oscillations \cite{Il'ichev1} can be classified as a weak continuous quantum measurement.

\section {Conclusions.}

In this paper we have analyzed quantitatively a continuous spectroscopic measurement of Rabi oscillations in a flux qubit by means of a high quality tank (LC circuit) which is inductively coupled to the qubit loop.
This circuit serves as a linear detector for measuring the spectrum of voltage fluctuations in the tank as well as  for monitoring the effective impedance of the system "qubit+tank". Expressions for the voltage spectrum and for the angle between the current that drives the tank and the averaged tank voltage have been derived in terms of the spectrum of qubit fluctuations and the qubit magnetic susceptibility. To find the spectrum of the qubit and its magnetic response we have applied a formalism of non-Markovian Heisenberg-Langevin equations to the case of a strongly driven open quantum system. Combining the Bloch-Redfield and rotating wave approximations
we have obtained formulas for the damping rates of the qubit and its magnetic susceptibility as functions of the amplitude of the high-frequency driving field and detuning of this field from the qubit energy splitting.
A dissipative evolution of the averaged current in the qubit loop and the probability to find the driven qubit in the excited state have been described analytically. Contributions of the qubit to the damping rate and the frequency shift of the tank have been calculated. We have presented also analytical formulas for the nonequilibrium spectrum of current fluctuations in the qubit loop as well as for the spectrum of voltage fluctuations in the tank (detector) which contains information about the Rabi frequency $\Omega_R$ and about the decay rate of Rabi oscillations $\Gamma$. It is shown that the ratio between the qubit contribution to the spectrum of voltage fluctuations and the thermal spectrum of the tank is peaked when the Rabi frequency is about the resonant frequency of the tank $\omega_T$. It corresponds to the maximal acquisition of information from the qubit. We have shown also that this effective measurement is accompanied by the maximal value of decoherence resulting from the backaction of the tank on the qubit.  The signal-to-noise ratio as a function of a deviation between the Rabi frequency and the frequency of the tank has a linewidth that is proportional to the qubit decay rate $\Gamma$, whereas measurement-induced decoherence of the qubit as a function of the same deviation is determined by the linewidth of the tank $\gamma_T$ which is much less than $\Gamma$. It allows us to find optimal conditions for the efficient spectroscopic measurement of the Rabi oscillations in the strongly driven quantum bit.

\begin{acknowledgements}
The author is grateful to Mohammad Amin, Alec Maassen van den Brink,  Miroslav Grajcar, Evgeni Il'ichev, Andrei Izmalkov, Nikolai Oukhanski, Alexander Shnirman, and Alexandre Zagoskin for many enlightening discussions. It is my pleasure also to thank Jeremy Hilton and Alexandre Zagoskin for critical reading of the manuscript.
\end{acknowledgements}

\appendix
\section{ Method of quantum Langevin equations.}
In this Appendix we sketch out basics of our approach to the theory of open quantum systems which has been proposed in Ref.\cite{Efremov1} and developed in Refs. \cite{Smirnov2,Mourokh1}. Formulas for the collision terms $L_x, L_y,L_z $ (29) will be derived in the process. Besides that, we give here explicit expressions for the fluctuation sources $\xi_x,\xi_y,\xi_z$ together with a recipe for calculating their correlation functions.

The Heisenberg equations (23) incorporate the total heat bath operator $Q(t)$ (24) multiplied by an operator of the qubit, say,  $   {\cal {A}} {(t)}  . $ These operators commute because they belong to the different physical systems. It is convenient to work with the symmetrized product of these operators. With the expansion (24) in mind we obtain
\begin{equation}
{1\over 2} [ Q(t), {\cal {A}} {(t)}]_+ = {1\over 2} [ Q^{(0)}(t), {\cal {A}} {(t)}]_+
+ \int dt_1 \varphi (t,t_1) {1\over 2} [ \sigma_z(t_1), {\cal {A}} {(t)}]_+ .
\end{equation}
The averaged value of the first parametric term in this expression is determined by the
quantum Furutsu-Novikov theorem \cite{Efremov1}:
\begin{equation}
\langle {1\over 2} [ Q^{(0)}(t), {\cal {A}} {(t)}]_+\rangle  =
\int dt' M(t,t')
\langle {\delta { \cal {A}} {(t)}  \over  \delta Q^{(0)}(t')  } \rangle ,
\end{equation}
where $M(t,t')$ is the symmetrized correlator of unperturbed heat bath variables,
$ M(t,t') =\langle (1/2) [Q^{(0)}(t),Q^{(0)}(t')]_+\rangle.$
This relation follows from the fact that due to the qubit-bath interaction a Heisenberg operator of the qubit represents a functional of the bath variables $ \{Q^{(0)}\}: {\cal {A}} {(t)} = {\cal {A}} { [ \{Q^{(0)}\},(t)]}$ that can be expanded in a sum of various time-ordered products like $  Q^{(0)}(t_{\alpha_1 })  Q^{(0)}(t_{\alpha_2}) ... Q^{(0)}(t_{\alpha_n}).$ The operators of the free heat bath, $ Q^{(0)},$  obey the Wick theorem. Because of this  the average value of the additional operator
$ Q^{(0)}(t) $ multiplied by the term $  Q^{(0)}(t_{\alpha_1 })  Q^{(0)}(t_{\alpha_2}) ... Q^{(0)}(t_{\alpha_n}) $   is reduced to the sum of pairings between the external operator  $ Q^{(0)}(t) $ and each operators  from the above-mentioned product:
$$
\langle Q^{(0)}(t) \cdot  Q^{(0)}(t_{\alpha_1} )  Q^{(0)}( t_{\alpha_2} ) ... Q^{(0)}(t_{\alpha_n} ) \rangle = \langle Q^{(0)}(t) Q^{(0)}(t_{\alpha_1} )\rangle \langle  Q^{(0)}( t_{\alpha_2} ) ... Q^{(0)}( t_{\alpha_n} ) \rangle + $$
$$ \langle Q^{(0)}(t) Q^{(0)}( t_{\alpha_2} )\rangle \langle  Q^{(0)}( t_{\alpha_1} ) Q^{(0)}( t_{\alpha_3} ) ... Q^{(0)}( t_{\alpha_n} ) \rangle + ... $$ $$
\langle Q^{(0)}(t) Q^{(0)}( t_{\alpha_n } )\rangle \langle  Q^{(0)}( t_{\alpha_1} ) Q^{(0)}( t_{\alpha_2} ) ... Q^{(0)}( t_{\alpha_{n-1}} ) \rangle. $$
The operator $Q^{(0)}(t_{\alpha_k }), k = 1,..,n, $ engaged with the operator
$ Q^{(0)}(t) $ in the outside correlator $ \langle Q^{(0)}(t) Q^{(0)}(t_{\alpha_k })\rangle $, should be removed from the initial product. It corresponds to taking a functional derivative over this variable:
$$
\langle Q^{(0)}(t) \cdot  Q^{(0)}(t_{\alpha_1 })  Q^{(0)}(t_{\alpha_2}) ... Q^{(0)}(t_{\alpha_n}) \rangle = $$ $$ \int dt' \langle Q^{(0)}(t) Q^{(0)}(t')  \rangle
\langle {\delta \over \delta Q^{(0)}(t')}
\{ Q^{(0)}(t_{\alpha_1 })  Q^{(0)}(t_{\alpha_2}) ... Q^{(0)}(t_{\alpha_n}) \} \rangle.
$$
This equation together with a relation $\delta Q^{(0)}(t_{\alpha})/\delta  Q^{(0)}(t')  = \delta(t_{\alpha} - t') $ results in the formula:
\begin{equation}
\langle Q^{(0)}(t) {\cal {A}} {(t)} \rangle =
 \int dt' \langle Q^{(0)}(t) Q^{(0)}(t')  \rangle  \langle {\delta { \cal {A}} {(t)}  \over  \delta Q^{(0)}(t')  } \rangle .
\end{equation}
We notice that the position of the external operator  $ Q^{(0)}(t) $ with respect to the operator $ {\cal {A}} {(t)} $ is mapped into a relative order of operators in the commutator $ \langle Q^{(0)}(t) Q^{(0)}(t')  \rangle  $ involved in Eq. (A3); in so doing the average value of the symmetrized product of $  Q^{(0)}(t) $ and $ {\cal {A}} {(t)} $ (see Eq.(A1)) is determined by the symmetrized correlator of the heat bath $M(t,t')$.

The functional derivative over the variable $Q^{(0)}(t')$ is equivalent to the derivative over the deterministic force $f(t')$ which is additive to  $Q^{(0)}(t')$  in the Hamiltonian (1). In its turn, the functional derivative of the qubit operator ${\cal {A}} {(t)}  $ over the force $f(t')$ is proportional to the commutator of  ${\cal {A}} {(t)}  $ and the qubit matrix $\sigma_z(t')$ that is conjugated to the force $f(t')$  in Eq.(1):
\begin{equation}
{\delta { \cal {A}} {(t)}  \over \delta Q^{(0)}(t')} = {\delta  {\cal {A}} {(t)} \over \delta f(t')} = i [ {\cal {A}} {(t)}, \sigma_z(t')]_- \theta (t-t').
\end{equation}
To show this we consider the Heisenberg operator of the qubit, ${\cal {A}} {(t)},$   in the interaction representation, when the interaction between the qubit and the force $f(t)$ is described by the term $H_{int} = - \sigma_z f(t).$ Then, an evolution of the operator $ {\cal {A}} {(t)},$
\begin{equation}
{\cal {A}} {(t)}  = S^+(t) { \cal {A}}^{(0)} {(t)}   S(t),
\end{equation}
is determined by the S-matrix, $S(t) = S(t, -\infty ),$ with
\begin{equation}
S(t,t_0) = T \{ \exp[i \int_{t_0}^{t} dt_1 \sigma_z^{(0)}(t_1) f(t_1)]  \}.
\end{equation}
Here $T$ is a time-ordering operator, $\sigma_z^{(0)}(t), {\cal {A}}^{(0)} {(t)}   $ are the free qubit operators (without the interaction with the force $f(t)$ ).
Then, for the functional derivative we obtain
$$
{\delta {\cal{A}} (t) \over \delta f(t') } = {\delta S^+(t) \over \delta f(t')}
{ \cal {A}}^{(0)} {(t)} S(t) + S^+(t) { \cal {A}}^{(0)} {(t)} {\delta S(t) \over \delta f(t')},$$
where
$$
{\delta S(t) \over \delta f(t')} = i \theta (t - t') T\{ \sigma_z^{(0)}(t')
\exp[i \int_{-\infty}^{t} dt_1 \sigma_z^{(0)}(t_1) f(t_1)] \} =  $$ $$
i \theta (t - t') T\{ \sigma_z^{(0)}(t')
\exp[i \int_{-\infty}^{t'} dt_1 \sigma_z^{(0)}(t_1) f(t_1)]
\exp[i \int_{t'}^{t} dt_1 \sigma_z^{(0)}(t_1) f(t_1)]\} = $$ $$
i \theta (t - t') T\{\exp[i \int_{t'}^{t} dt_1 \sigma_z^{(0)}(t_1) f(t_1)]\}
\sigma_z^{(0)}(t') T\{ \exp[i \int_{-\infty}^{t'} dt_1 \sigma_z^{(0)}(t_1) f(t_1)] \}
= $$ $$ i \theta (t - t') S(t,t') \sigma_z^{(0)}(t') S(t')
= i \theta (t - t') S(t) S^{+}(t') \sigma_z^{(0)}(t') S(t') = i \theta (t - t') S(t)
\sigma_z(t'), $$
and $\sigma_z(t') $ is the total Heisenberg operator.
Here we apply the relation $S(t,t') = S(t) S^{-1}(t') = S(t)S^{+}(t'),$ which  follows from the facts that $S^{-1}(t') = S^+(t'),$ and
$$ S(t,t') S(t') = S(t,t') S(t',-\infty) = S(t,-\infty) = S(t). $$
Taking into account the derivative of the matrix $S^{+}: \delta S^{+}(t)/\delta f(t') =
- i \theta (t - t') \sigma_z(t') S^{+}(t), $ we obtain Eq.(A4)  for the functional derivative of an arbitrary Heisenberg operator $ {\cal {A}} {(t)}.$

In view of the Furutsu-Novikov theorem (A2) the operators  like $(1/2) [ Q(t), {\cal {A}} {(t)}]_+ $ involved in the Heisenberg equations  (23) can be splitted into a fluctuation force $ \xi_{\cal {A} }  $ and a collision term $ L_ {\cal {A}}$ :
\begin{equation}
  (1/2) [ Q(t), {\cal {A}} {(t)}]_+ = \xi_{\cal {A} }  + L_ {\cal {A}},
\end{equation}
where  the fluctuation force,
\begin{equation}
\xi_{\cal {A} } (t)  = \{ Q^{(0)}(t), {\cal {A}} {(t)} \} =
{1\over 2} [ Q^{(0)}(t), {\cal {A}} {(t)}]_+  - \int dt' M(t,t')
 {\delta { \cal {A}} {(t)}  \over  \delta f(t')  },
\end{equation}
has a zero average value, $\langle \xi_{\cal {A} }  \rangle = 0, $
 and the collision term,
\begin{equation}
L_ {\cal {A}}(t) =  \int dt_1 \tilde{M}(t,t_1)i [ {\cal {A}} {(t)}, \sigma_z(t_1)]_-  + \int dt_1 \varphi (t,t_1) {1\over 2}[ {\cal {A}} {(t)},\sigma_z(t_1)]_{+},
\end{equation}
incorporates contributions  both parametric fluctuations and a back action of the heat bath. Here we introduce a causal correlation function of the free heat bath $\tilde{M}(t,t_1) = M(t,t_1) \theta(t-t_1)$ having $ \tilde{S}(\omega ) $ as a Fourier transform,
\begin{equation}
 \tilde{S}(\omega ) =  \int d\tau e^{i\omega \tau } \tilde{M}(\tau) = \int {d\omega_1 \over 2 \pi}{i \over \omega - \omega_1 + i\epsilon}S(\omega_1),
 \end{equation}
with  $S(\omega)$ being the equilibrium spectrum of the heat bath (27), and $\epsilon \rightarrow +0.$

The explicit form of the fluctuation sources allows us to find their correlation functions. To do that we have to take pairings of all free heat bath variables $Q^{(0)}$ with the heat bath variables and the qubit operators belonging to   other fluctuation forces. In the case of weak qubit-bath coupling we can take into account pairings between the free heat bath variables only. With this procedure  we derive the following expressions for a  correlator of fluctuation forces $\xi_A(t)$ and $\xi_B(t)$: $
\langle \xi_A(t) \xi_B(t') \rangle = \langle Q^{(0)}(t), Q^{(0)}(t')\rangle
\langle {\cal {A}} {(t)} {\cal {B}} {(t')} \rangle, $ and  for the symmetrized correlation function:
\begin{equation}
\langle {1\over 2}\left[\xi_A(t), \xi_B(t')\right]_+ \rangle =
M(t,t')  \langle {1\over 2}\left[  {\cal {A}} {(t)} , {\cal {B}} {(t')}  \right]_+ \rangle + R(t,t')  \langle {1\over 2}\left[  {\cal {A}} {(t)} , {\cal {B}} {(t')}  \right]_- \rangle
\end{equation}
where the antisymmetrized correlator of the heat bath, $R(t,t') = \langle (1/2) [Q^{(0)}(t),Q^{(0)}(t')]_-\rangle$, has the spectral function $\chi^{\prime\prime}(\omega)$ as its Fourier transform.

\section{Collision integrals.}

Collision terms $L_x,L_y,L_z$ (29) can be simplified in the approximation of weak coupling between the qubit and the heat bath. In this case (anti)commutators of the qubit variables $X,Y,Z$ taken at different moments of time
are calculated with free evolution operators of the qubit (31). Here we present expressions for
(anti)commutators of qubit operators $X,Y,Z$ involved both into the collision terms and into the correlation functions of the fluctuation forces.
With Eqs. (31) and the usual commutation rules we obtain ( here  $\tau = t - t_1 $):
\begin{eqnarray}
i [X(t),X(t_1)]_- = 2 Z(t_1) {A\over \Omega_R} \sin \Omega_R \tau - 2 Y(t_1)
{\delta A \over \Omega_R^2} (1 - \cos \Omega_R \tau), \nonumber\\
{1\over 2}[X(t),X(t_1)]_+ = {\delta^2 \over \Omega_R^2} + {A^2 \over \Omega_R^2} \cos \Omega_R\tau ; \nonumber\\
i [X(t),Y(t_1)]_-  = - 2 Z(t_1) \left( {\delta^2 \over \Omega_R^2} + {A^2 \over \Omega_R^2} \cos \Omega_R\tau \right)+ 2 X(t_1)  {\delta A \over \Omega_R^2} (1 - \cos \Omega_R \tau), \nonumber\\
{1\over 2}[X(t),Y(t_1)]_+ = -{1\over 2}[Y(t),X(t_1)]_+  = {A\over \Omega_R} \sin \Omega_R \tau ; \nonumber\\
i [Y(t),X(t_1)]_- = 2 Z(t_1) \cos \Omega_R\tau - 2 Y(t_1) {\delta  \over \Omega_R}
\sin \Omega_R \tau ; \nonumber\\
i [X(t),Z(t_1)]_- =  2 Y(t_1) \left( {\delta^2 \over \Omega_R^2} + {A^2 \over \Omega_R^2} \cos \Omega_R\tau \right) - 2 X(t_1) {A\over \Omega_R} \sin \Omega_R \tau , \nonumber\\
{1\over 2}[X(t),Z(t_1)]_+ = {1\over 2}[Z(t),X(t_1)]_+  =  {\delta A \over \Omega_R^2} (1 - \cos\Omega_R \tau), \nonumber\\
i [Z(t),X(t_1)]_- =  - 2 Y(t_1) \left( {A^2 \over \Omega_R^2} + {\delta^2 \over \Omega_R^2} \cos \Omega_R\tau \right) - 2 Z(t_1) {\delta \over \Omega_R}\sin \Omega_R \tau ; \nonumber\\
i [Y(t),Y(t_1)]_- =  2 Z(t_1) {A\over \Omega_R} \sin \Omega_R \tau + 2 X(t_1)
 {\delta \over \Omega_R} \sin \Omega_R \tau , \nonumber\\
{1\over 2}[Y(t),Y(t_1)]_+ = \cos\Omega_R \tau; \nonumber\\
i [Y(t),Z(t_1)]_-  = - 2 X(t_1)  \cos\Omega_R \tau - 2 Y(t_1) {A\over \Omega_R} \sin \Omega_R \tau , \nonumber\\
{1\over 2}[Y(t),Z(t_1)]_+  = - {1\over 2}[Z(t),Y(t_1)]_+ =
{\delta  \over \Omega_R} \sin \Omega_R \tau ; \nonumber\\
i [Z(t),Y(t_1)]_- =  2 X(t_1) \left( {A^2 \over \Omega_R^2} + {\delta^2 \over \Omega_R^2} \cos \Omega_R\tau \right) -  2 Z(t_1) {\delta A \over \Omega_R^2} (1 - \cos \Omega_R \tau); \nonumber\\
i [Z(t),Z(t_1)]_-  = 2 X(t_1) {\delta  \over \Omega_R} \sin \Omega_R \tau +
 2 Y(t_1) {\delta A \over \Omega_R^2} (1 - \cos \Omega_R \tau), \nonumber\\
{1\over 2}[Z(t),Z(t_1)]_+ =  \left( {A^2 \over \Omega_R^2} + {\delta^2 \over \Omega_R^2} \cos \Omega_R\tau \right).
\end{eqnarray}

Using these formulas we find the following expressions for the collision coefficients involved in Eq.(33):
\begin{eqnarray}
\bar{\Gamma}_{xx} (\tau ) = 2{\Delta^2 \over \omega_c^2} \tilde{M}(\tau ) \left[
\left( 1 + {\delta^2\over \Omega_R^2} \right) \cos \Omega_R\tau  \cos\omega_0\tau + 2 {\delta \over \Omega_R} \sin\omega_0\tau \sin \Omega_R\tau  + {A^2 \over \Omega_R^2} \cos \omega_0\tau \right], \nonumber\\
\bar{\Gamma}_{xy} (\tau ) = - \bar{\Gamma}_{yx} (\tau ) =
2{\Delta^2 \over \omega_c^2} {A\over \Omega_R} \tilde{M}(\tau ) \left[ \sin\Omega_R\tau  \cos\omega_0\tau  + {\delta \over \Omega_R} ( 1 - \cos \Omega_R \tau ) \sin\omega_0\tau  \right], \nonumber\\
\bar{\Gamma}_{xz} (\tau ) = \bar{\Gamma}_{zx} (\tau ) =
2{\Delta^2 \over \omega_c^2} {A\over \Omega_R} \tilde{M}(\tau ) \left[ \sin\Omega_R\tau  \sin\omega_0\tau  - {\delta \over \Omega_R} ( 1 - \cos \Omega_R \tau ) \cos\omega_0\tau  \right], \nonumber\\
\bar{\Gamma}_{yy} (\tau ) = 2 \tilde{M}(\tau ) \left[ 2 {\varepsilon^2 \over \omega_c^2} \left( {A^2 \over \Omega_R^2} + {\delta^2 \over \Omega_R^2} \cos \Omega_R\tau \right) + {\Delta^2 \over \omega_c^2} \cos \omega_0 \tau
\left( {\delta^2 \over \Omega_R^2} + {A^2 \over \Omega_R^2} \cos \Omega_R\tau \right)
\right], \nonumber\\
\bar{\Gamma}_{yz} (\tau ) = - \bar{\Gamma}_{zy} (\tau ) =
2 \tilde{M}(\tau )\left[ 2 {\varepsilon^2 \over \omega_c^2} {\delta \over \Omega_R} \sin\Omega_R\tau  +  {\Delta^2 \over \omega_c^2} \sin \omega_0 \tau
\left( {\delta^2 \over \Omega_R^2} + {A^2 \over \Omega_R^2} \cos \Omega_R\tau \right)
\right], \nonumber\\
\bar{\Gamma}_{zz} (\tau ) = 2 \tilde{M}(\tau )
\left[ 2 {\varepsilon^2 \over \omega_c^2}  \cos \Omega_R\tau  + {\Delta^2 \over \omega_c^2} \cos \omega_0 \tau
\left( {\delta^2 \over \Omega_R^2} + {A^2 \over \Omega_R^2} \cos \Omega_R\tau \right)
\right].
\end{eqnarray}
Fourier transforms of the collision coefficients $ \bar{\Gamma}_{mn} (\tau )$ (B2),
$$
\tilde{\Gamma}_{mn}(\omega ) = \int d\tau e^{i \omega \tau} \bar{\Gamma}_{mn} (\tau ),
$$
 are proportional to the causal spectrum of the heat bath $\tilde{S}(\omega )$ (A10).
Frequency-dependent relaxation rates $\Gamma_z(\omega ), \Gamma (\omega )$  (39),(40) defining relaxation and decoherence of the qubit are expressed in terms of real and imaginary parts of the functions
$\tilde{\Gamma}_{mn}(\omega ) :$
\begin{eqnarray}
\Gamma_z(\omega ) = {\delta^2 \over \Omega_R^2} \tilde{\Gamma}_{xx}^{\prime}(\omega ) + 2 {\delta A \over \Omega_R^2}
\tilde{\Gamma}_{xz}^{\prime}(\omega ) +  {A^2 \over \Omega_R^2} \tilde{\Gamma}_{zz}^{\prime}(\omega )  , \nonumber\\
 \Gamma(\omega ) = {A^2 \over \Omega_R^2} \tilde{\Gamma}_{xx}^{\prime}(\omega )  +  \tilde{\Gamma}_{yy}^{\prime}(\omega ) +
{\delta^2 \over \Omega_R^2} \tilde{\Gamma}_{zz}^{\prime}(\omega )  + \nonumber\\
2 {\delta \over \omega }  \tilde{\Gamma}_{yz}^{\prime\prime}(\omega ) +
2 { A \over \omega }  \tilde{\Gamma}_{xy}^{\prime\prime}(\omega ) -
2 {\delta A \over \Omega_R^2} \tilde{\Gamma}_{xz}^{\prime}(\omega ).
\end{eqnarray}

\section{Correlators of fluctuation forces.}

In this appendix we adduce formulas for the spectral functions of fluctuation forces $K_{nl}(\omega )$
that eventually determine the nonequilibrium spectrum of qubit fluctuations $S_{zz}(\omega )$ (57).
Correlation functions of fluctuation forces, $ \xi_m(t) = \{ Q^{(0)}(t), {\cal {A}}_m {(t)} \}, $
 are calculated according to Eq.(A9) with (anti)commutators  presented by Eqs.(B1).
For the spectrum $K_{yy}(\omega ),$ as an example, the corresponding correlator of the fluctuation forces,
$ \langle (1/2)\left[\xi_y(t), \xi_y(t')\right]_+ \rangle $ is obtained from Eq.(A9) with the operators ${\cal {A}} =   {\cal {A}}_y {(t)}, {\cal {B}} {(t')} = {\cal {A}}_y {(t')}$ (30):
\begin{eqnarray}
\langle {1\over 2}\left[\xi_y(t), \xi_y(t')\right]_+ \rangle = \nonumber\\
2{\Delta^2 \over \omega_c^2} \left\{
M(t,t') \langle {1\over 2}\left[X(t), X(t')\right]_+ \rangle +
R(t,t') \langle {1\over 2}\left[X(t), X(t')\right]_- \rangle \right\} \cos\omega_0(t-t') + \nonumber\\
4 {\varepsilon^2 \over \omega_c^2}
\left\{
M(t,t') \langle {1\over 2}\left[Z(t), Z(t')\right]_+ \rangle +
R(t,t') \langle {1\over 2}\left[Z(t), Z(t')\right]_- \rangle \right\}.
\end{eqnarray}
Following to this procedure for all correlation functions of the fluctuation forces and taking corresponding Fourier transforms we find for the spectral functions of fluctuation forces
$K_{lq}(\omega ); l,q = x,y,z:$
\begin{eqnarray}
K_{xx}(\omega ) = {\Delta^2 \over  \omega_c^2} [ S(\omega + \omega_0) + S(\omega - \omega_0)] + \nonumber\\
{\Delta^2 \over 2 \omega_c^2}
[ S(\omega + \omega_0 + \Omega_R) + S(\omega - \omega_0 - \Omega_R ) +  \nonumber\\
S(\omega + \omega_0 - \Omega_R) + S(\omega - \omega_0 + \Omega_R )]  +   \nonumber\\
{\Delta^2 \over 2 \omega_c^2}
P_0 [ \chi^{\prime\prime}(\omega - \omega_0 - \Omega_R) - \chi^{\prime\prime}(\omega + \omega_0 + \Omega_R) +  \nonumber\\
\chi^{\prime\prime}(\omega + \omega_0 - \Omega_R) -
 \chi^{\prime\prime}(\omega - \omega_0 + \Omega_R)] ;
\end{eqnarray}
\begin{eqnarray}
K_{xy}(\omega ) = - K_{yx}(\omega ) =
 - i  {\Delta^2 \over 2 \omega_c^2} {A\over \Omega_R}
[ S(\omega + \omega_0 + \Omega_R) - S(\omega - \omega_0 - \Omega_R ) + \nonumber\\
 S(\omega - \omega_0 + \Omega_R) - S(\omega + \omega_0 - \Omega_R )] +
\nonumber\\
i  {\Delta^2 \over 2 \omega_c^2} {A\over \Omega_R}
P_0 [ \chi^{\prime\prime}(\omega + \omega_0 +\Omega_R) + \chi^{\prime\prime}(\omega - \omega_0 - \Omega_R) + \nonumber\\
\chi^{\prime\prime}(\omega + \omega_0 - \Omega_R) +
 \chi^{\prime\prime}(\omega - \omega_0 + \Omega_R)];
\end{eqnarray}
\begin{eqnarray}
K_{xz}(\omega ) = K_{zx}(\omega ) =
{\Delta^2 \over 2 \omega_c^2} {A\over \Omega_R}
[ S(\omega + \omega_0 - \Omega_R) + S(\omega - \omega_0 + \Omega_R ) - \nonumber\\
S(\omega + \omega_0 + \Omega_R) - S(\omega - \omega_0 - \Omega_R )] + \nonumber\\
{\Delta^2 \over 2 \omega_c^2} {A\over \Omega_R} P_0 [ \chi^{\prime\prime}(\omega + \omega_0 - \Omega_R) - \chi^{\prime\prime}(\omega - \omega_0 + \Omega_R) + \nonumber\\
 \chi^{\prime\prime}(\omega + \omega_0 + \Omega_R) -  \chi^{\prime\prime}(\omega - \omega_0 - \Omega_R)];
\end{eqnarray}
\begin{eqnarray}
K_{yy}(\omega ) =  4 {\varepsilon^2\over \omega_c^2} S(\omega ) +
{\Delta^2 \over 2 \omega_c^2}
[ S(\omega + \omega_0 + \Omega_R) + S(\omega - \omega_0 - \Omega_R ) + \nonumber\\
 S(\omega + \omega_0 - \Omega_R) + S(\omega - \omega_0 + \Omega_R )]  +  \nonumber\\
{\Delta^2 \over 2 \omega_c^2}
P_0 [ \chi^{\prime\prime}(\omega - \omega_0 - \Omega_R) - \chi^{\prime\prime}(\omega + \omega_0 + \Omega_R) + \nonumber\\
\chi^{\prime\prime}(\omega + \omega_0 - \Omega_R) -
 \chi^{\prime\prime}(\omega - \omega_0 + \Omega_R)];
\end{eqnarray}
\begin{eqnarray}
K_{yz}(\omega ) = K_{zy}(\omega ) =
-i {\Delta^2 \over 2 \omega_c^2} [ S(\omega + \omega_0 + \Omega_R) - S(\omega - \omega_0 - \Omega_R ) + \nonumber\\
 S(\omega + \omega_0 - \Omega_R) - S(\omega - \omega_0 + \Omega_R )] - \nonumber\\
i {\Delta^2 \over 2 \omega_c^2} P_0 [ \chi^{\prime\prime}(\omega + \omega_0 - \Omega_R) + \chi^{\prime\prime}(\omega - \omega_0 + \Omega_R) - \nonumber\\
 \chi^{\prime\prime}(\omega + \omega_0 + \Omega_R) -  \chi^{\prime\prime}(\omega - \omega_0 - \Omega_R)];
\end{eqnarray}
\begin{eqnarray}
K_{zz}(\omega ) =  2{\varepsilon^2\over \omega_c^2} [ S(\omega +\Omega_R ) + S(\omega -\Omega_R ) - \nonumber\\
P_0 \chi^{\prime\prime}(\omega +  \Omega_R)  +   P_0 \chi^{\prime\prime}(\omega -  \Omega_R)] + \nonumber\\
{\Delta^2 \over 2 \omega_c^2}
[ S(\omega + \omega_0 + \Omega_R) + S(\omega - \omega_0 - \Omega_R ) + \nonumber\\
S(\omega + \omega_0 - \Omega_R) + S(\omega - \omega_0 + \Omega_R )]  +  \nonumber\\
{\Delta^2 \over 2 \omega_c^2}
P_0 [ \chi^{\prime\prime}(\omega - \omega_0 - \Omega_R) - \chi^{\prime\prime}(\omega + \omega_0 + \Omega_R) + \nonumber\\
\chi^{\prime\prime}(\omega + \omega_0 - \Omega_R) -
 \chi^{\prime\prime}(\omega - \omega_0 + \Omega_R)].
\end{eqnarray}


\begin{references}
\bibitem{footnote1}
Signs should be reversed before the parentheses in the second and third equations (5) of Ref.\cite{Smirnov1}.
\end{references}
\end{document}